\documentclass[11pt,nofootinbib,prd,showpacs,notitlepage,tightenlines,longbibliography]{revtex4-1}

\usepackage{amsmath}
\usepackage{amssymb}
\usepackage{stmaryrd}
\usepackage{dsfont}  
\usepackage{bbm}
\usepackage{graphicx}

\newcommand{\ket}[1]{| #1 \rangle}
\newcommand{\bra}[1]{\langle #1 |}
\newcommand{\scalar}[2]{\left\langle #1 | #2 \right\rangle}
\newcommand{\mean}[1]{\langle #1 \rangle}

\newcommand{\be}{\nopagebreak[3]\begin{equation}}
\newcommand{\ee}{\end{equation}}

\newtheorem{prop}{Proposition}

\newcommand{\normord}[1]{:\mathrel{\mspace{2mu}#1\mspace{2mu}}:}

\begin{document}

\title{Loop expansion and the bosonic representation of loop quantum gravity}
\author{E. Bianchi}
\email{ebianchi@gravity.psu.edu}
\author{J. Guglielmon}
\email{jag585@psu.edu}
\author{L. Hackl}
\email{lucas.hackl@psu.edu}
\author{N. Yokomizo}
\email{yokomizo@gravity.psu.edu}
\affiliation{Institute for Gravitation and the Cosmos \& Physics Department,\\ Penn State, University Park, PA 16802, USA}
\begin{abstract}
We introduce a new loop expansion that provides a resolution of the identity in the Hilbert space of loop quantum gravity on a fixed graph. We work in the bosonic representation obtained by the canonical quantization of the spinorial formalism. The resolution of the identity gives a tool for implementing the projection of states in the full bosonic representation onto the space of solutions to the Gauss and area matching constraints of loop quantum gravity. This procedure is particularly efficient in the semiclassical regime, leading to explicit expressions for the loop expansions of coherent, heat kernel and squeezed states.

\end{abstract}
\date{\today}
\maketitle

\section{Introduction}

The most characteristic feature of the loop approach to quantum gravity is the representation of non-perturbative states of the quantized gravitational field in terms of extended excitations with support on closed loops \cite{Rovelli:2004tv,Thiemann:2007zz,Gambini:1996ik,Ashtekar:2004eh}. A loop state $\ket{\alpha}$ corresponds to an elementary quantum excitation of a single Faraday line of the gravitational field as described by the Ashtekar-Barbero connection $A_a^i(x)$. By construction, any loop state is gauge-invariant, and diffeomorphism invariance is implemented by letting the $\alpha$ be s-knots, i.e., isotopy classes of loops. An infinite class of solutions to the Hamiltonian constraint was found in this approach, leading for the first time to the construction of exact solutions to the full set of constraints of canonical quantum gravity \cite{Jacobson:1987qk,Rovelli:1989za}. This result was the main motivation underlying the early stages of development of loop quantum gravity. 

A loop state $\ket{\alpha}$ is defined by the action of the corresponding Wilson loop operator $W_\alpha$ on the vacuum of the theory, $\ket{\alpha}=W_\alpha \ket{0}$. The definition naturally extends to multiloops $\Phi=\{\alpha_i\}$ by setting $W_{\Phi}=\prod_i  W_{\alpha_i}$. The kinematical Hilbert space $\mathcal{H}$ of loop quantum gravity is spanned by such loop states, allowing arbitrary states to be written as superpositions of multiloop excitations,
\be
\ket{\psi} = \sum_\Phi  c_\Phi \, W_\Phi \ket{0} \, .
\label{eq:wilson-loop-expansion}
\ee
However, since Wilson loops are related by the Mandelstam and retracing identities \cite{Gambini:1996ik,Rovelli:1995ac}, multiloop states are not independent, satisfying in fact a large number of nonlocal identities. As a result, the loop basis is highly overcomplete, leading to severe technical difficulties in dealing with states in the form \eqref{eq:wilson-loop-expansion}. The standard solution to this problem consists in expanding states in the spin network basis instead, an orthonormal basis formed by linear combinations of loop states that completely reduce the Mandelstam identities \cite{Rovelli:1995ac,Baez:1994hx,Baez:1995md}. In this paper, we introduce a new procedure that allows us to define a resolution of the identity in the loop representation that addresses the difficulties of the loop basis while retaining its physical properties.

The key technical tool underlying our construction is the spinor formalism of loop quantum gravity introduced in \cite{Girelli:2005ii} and further developed in several works \cite{Freidel:2010tt,Dupuis:2010iq,Borja:2010rc,Bonzom:2012bn,Livine:2011gp,Livine:2011zz,Livine:2013wmq}. In this formalism, states of loop quantum gravity on a fixed graph $\Gamma$ are reformulated in terms of bosonic variables $a_i,a_i^\dagger$ that essentially consist of an adaptation of the Schwinger oscillator model of angular momentum \cite{Schwinger:2015,Sakurai:2011zz} to the context of quantum gravity. Writing the Wilson loops $W_\Phi$ in terms of creation and annihilation operators \cite{Livine:2013wmq,Bianchi:2016}, one can speak of their normal ordered version $\normord{W_\Phi}$. We consider a new loop expansion of the form:
\be
\ket{\psi} = \sum_\Phi c_\Phi \normord{W_\Phi} \ket{0} \, .
\label{eq:F-loop-expansion}
\ee
Allowing $\Phi$ to take values in a space of non-repeating loops defined later in the paper, the loop states $\ket{\Phi}=\normord{W_\Phi} \ket{0}$ form a new basis of  the space of states on a graph $\Gamma$. 

In this picture, several redundancies present in the original Wilson loop expansion are automatically solved. The retracing identity is completely reduced: in the equivalence class of all loops related by the addition or removal of trivial segments of the form $\gamma\circ\gamma^{-1}$, only a single representative contributes to the expansion \eqref{eq:F-loop-expansion}. This in turn reduces dramatically the number of Mandelstam identities, since it then suffices to consider the local ones. The new loop basis is still overcomplete, but only \emph{local} Mandelstam identities that relate partially overlapping loops are present. As a result, a resolution of the identity for states with support on a graph $\Gamma$ can be derived:
\be
P_{\Gamma} = \sum_\Phi p(\Phi) \ket{\Phi} \bra{\Phi} \, ,
\label{eq:res-identity}
\ee
where $p(\Phi)$ is a simple combinatorial function of the multiloop $\Phi$. The resolution of the identity \eqref{eq:res-identity} allows the coefficients $c_\Phi$ in the loop expansion \eqref{eq:F-loop-expansion} to be determined whenever the scalar products $\scalar{\Phi}{\psi}$ can be computed. It turns out that this is the case for several known families of semiclassical states in loop quantum gravity, including coherent \cite{Dupuis:2010iq,Bonzom:2012bn}, squeezed \cite{Bianchi:2016} and heat kernel states \cite{Thiemann:2000bw,Thiemann:2002vj,Bianchi:2009ky,Bianchi:2010mw}. The loop expansion coefficients for these states are given by simple Gaussian integrals that can be computed in the Bargmann representation of the oscillator model.

In general, the choice of a particular basis in the Hilbert space of a quantum system is dictated by the physical problem at hand. The spin network basis makes the intrinsic geometry transparent by diagonalizing the area and volume operators of the elementary quanta of space \cite{Rovelli:1994ge}. The behavior of the Wilson loop operators describing the extrinsic geometry is obscured in this representation, however. Moreover, the Hamiltonian constraint has a complicated form in the spin network basis, leading to severe difficulties in the study of the dynamics in the canonical approach. These are compelling motivations for the exploration of alternative bases. The new loop basis is a natural choice for further investigations of the dynamics given the success in the construction of solutions to the Hamiltonian constraint in this basis. The loop basis is also expected to be applicable to problems related to the semiclassical limit of loop quantum gravity. Our approach is based on the bosonic formalism also employed for the construction of coherent and squeezed states. The definition of these semiclassical states involves a projection to the space of states of loop quantum gravity in the (larger) bosonic space, and the projection operator \eqref{eq:res-identity} is precisely the tool required for that purpose, providing the means for a concrete description of such states.

This paper is organized as follows. In Section \ref{sec:bosonic} we review the reformulation of loop quantum gravity in terms of bosonic variables and discuss the representation of the holonomy-flux algebra in this formalism. The loop expansion of the projector to the space of states of loop quantum gravity on a graph $\Gamma$ is derived in Section \ref{sec:loop-expansion}, and applied to coherent, squeezed and heat kernel states in Section \ref{sec:semiclassical}. A closely related presentation of squeezed vacua in terms of a generating function is introduced in Section \ref{sec:generating-function}. We summarize and discuss the main results of the paper in Section \ref{sec:conclusion}. Two appendices include proofs of auxiliary results stated in the main text.

\section{Bosonic representation of loop quantum gravity}
\label{sec:bosonic}

In this section we review the reformulation of loop quantum gravity in terms of bosonic variables and discuss the representation of the holonomy-flux algebra and projectors to the space of solutions to the Gauss and area matching constraints in this formalism.

\subsection{Seeds, graphs and loops}
A finite graph with $N$ nodes and $L$ links can be defined combinatorially as follows. Let $\mathcal{S}$ be an ordered set consisting of $2L$ elements,
\begin{equation}
\mathcal{S}=\{1,\ldots, 2L\}\,.
\end{equation}
We call its elements \emph{seeds} and denote them by an index $i=1,\ldots, 2L$. The set $\mathcal{S}$ of seeds can be decomposed as the disjoint union of $N$ subsets, i.e. 
\begin{equation}
\mathcal{N}=\{n_1,\ldots,n_N\}\, , \quad \textrm{with} \quad \bigcup\limits_{k=1}^{N} n_k = \mathcal{S} \, .
\end{equation}
The elements $n=\{i_1,\ldots,i_{|n|}\}$ of $\mathcal{N}$ are called \emph{nodes}, and the number $|n|$ of seeds in $n$ is called the \emph{valence} of the node. If two seeds $i,j$ belong to the same node we write $i\sim j$, and say that they form a \emph{wedge} $w=\{i,j\}$ at the node $n$. The set $\mathcal{S}$ can also be decomposed as the disjoint union of $L$ subsets containing two elements each, i.e. 
\begin{equation}
\mathcal{L}=\{\ell_1,\ldots,\ell_L\}\quad \textrm{with} \quad \bigcup\limits_{k=1}^{L} \ell_k = \mathcal{S}\,.
\end{equation}
The elements $\ell$ of $\mathcal{L}$ are called \emph{links} and consist of ordered pairs of seeds, $\ell=(i,j)$ with $i<j$. We call $i=s(\ell)$ the \emph{source} and $j=t(\ell)$ the \emph{target} of the link $\ell$. Given a link $\ell=(s,t)$, we also define the link with reversed orientation $\ell^{-1}=(t,s)$. An oriented \emph{graph} $\Gamma$ is defined by the ordered set of its seeds $\mathcal{S}$, together with the two decompositions $\mathcal{N}$ and $\mathcal{L}$.

The graph $\Gamma=(\mathcal{S},\mathcal{N},\mathcal{L})$ has $N$ nodes and $L$ links.  Given a graph $\Gamma$ we can introduce loops and multiloops. 
Consider a sequence $\{\ell_1^{\epsilon_1},\ldots, \ell_{|\alpha|}^{\epsilon_{|\alpha|}}\}$ of links $\ell^{\epsilon}$ with orientation $\epsilon=\pm1$ such that $t(\ell_k^{\epsilon_k})\sim s(\ell_{k+1}^{\epsilon_{k+1}})$  and $t(\ell_{|\alpha|}^{\epsilon_{|\alpha|}})\sim s(\ell_{1}^{\epsilon_1})$. A \emph{loop} $\alpha=\{\ell_1^{\epsilon_1},\ldots, \ell_{|\alpha|}^{\epsilon_{|\alpha|}}\}$ is one such sequence up to cyclic permutations and up to an overall change of orientation. If there is no proper cyclic permutation that leaves the sequence invariant, we say that the loop is non-repeating. A loop can also be understood as a sequence of wedges $w=\{i,j\}$, i.e., couples of seeds at a node, so that $\alpha=\{w_1,\ldots,w_{|\alpha|}\}$. An oriented loop is a loop equipped with a choice of overall orientation.

A \emph{multiloop} $\Phi$ is a multiset formed by loops $\alpha$ with multiplicity $m_\alpha \in \mathbb{Z}^+$, $\Phi=\{{\alpha_1}^{m_1},{\alpha_2}^{m_2},\ldots\}$. A multiloop is non-repeating if it contains only non-repeating loops. If we flatten the multiloop $\Phi$\footnote{By flattening a collection of multisets $\{M_i\}$ we mean forming the union $M=\bigcup M_i$  and assigning as the multiplicity of each element $m \in M$ the sum of its multiplicities in each $M_i$.}, with loops written as sequences of links, we find that a link $\ell$ (up to orientation) can appear more than once, 
$\textrm{\emph{Flatten}}(\Phi)=\{\,\ell_1{}^{2j_1}\,,\ell_2{}^{2j_2}\,,\,\ldots\,\}$
where $2j_\ell$ is an integer.\footnote{The choice of notation in terms of a half-integer $j_\ell$ is meant to match the role of spin in spin-network states defined over the graph.} The half-integer $j_\ell=j_\ell(\Phi)$ is understood as a function of the multiloop. An oriented multiloop is a multiloop equipped with a choice of orientation for each of its loops.

\subsection{Bosonic lattice and holonomy-flux algebra}

We have defined a graph $\Gamma$ starting from the ordered set of seeds $\mathcal{S}$. Now we introduce a bosonic Hilbert space $\mathcal{H}_{\mathcal{S}}$ associated with $\mathcal{S}$, following standard techniques \cite{Girelli:2005ii,Livine:2011gp,Livine:2013wmq}. The Hilbert space of loop quantum gravity on the graph $\Gamma$ is a subspace of the bosonic Hilbert space, $\mathcal{H}_\Gamma\subset \mathcal{H}_{\mathcal{S}}$.\footnote{We restrict attention to the case of a fixed graph $\Gamma$. For an analysis of cylindrical consistency and an extension of the bosonic techniques to the continuum Hilbert space, see Section 4 of Ref. \cite{Livine:2011gp}}

To each seed $i$ in a graph $\Gamma$ we associate a pair of bosonic degrees of freedom labeled by an index $A=0,1$. As a result we have a bosonic system with $4L$ degrees of freedom, a \emph{bosonic lattice}. Creation and annihilation operators $a_i^A{}^\dagger, a_i^A$ satisfy the canonical commutation relations
\begin{equation}
[a_i^A,a_j^B{}^{\dagger}]=\,\delta_{ij}\,\delta^{AB},\qquad [a_i^A,a_j^B]=0,\qquad [a_i^A{}^{\dagger},a_j^B{}^{\dagger}]=0.
\end{equation}
The Hilbert space $\mathcal{H}_{\mathcal{S}}$ of the bosonic lattice is the Fock space built over the vacuum $|0\rangle$ defined as the state annihilated by all the operators $a_i^A$,
\begin{equation}
a_i^A|0\rangle=0\qquad \forall\;\,i=1,\ldots, 2L\,,\;\;A=0,1.
\end{equation}
A Hilbert subspace $\mathcal{H}_i$ generated by the action of the pair of creation operators $a_i^{\dagger A}$, $A=0,1$, on the vacuum $\ket{0}$ is naturally associated with each seed $i$. The full Hilbert space $\mathcal{H}_{\mathcal{S}}$ is the tensor product of all such seed subspaces, $\mathcal{H}_{\mathcal{S}}= \bigotimes_i \mathcal{H}_i$.

Creation and annihilation operators associated with wedges, loops and multiloops are defined in terms of the basic bosonic variables. For an oriented wedge $w=(i,j)$, we introduce the wedge annihilation operator:
\be
F_w = F_{ij} = \epsilon_{AB} a_i^A a_j^B \, ,
\label{eq:wedge-annihilation}
\ee
where $\epsilon_{AB}$ is the $2\times 2$ antisymmetric tensor with $\epsilon_{01}=+1$. If $i,j$ are seeds of distinct nodes, we set $F_{ij}=0$. In addition, for oriented loops $\alpha=\{w_1,\ldots,w_{|\alpha|}\}$ and oriented multiloops $\Phi=\{{\alpha_1}^{m_1},{\alpha_2}^{m_2},\ldots\}$, we define:
\be
F_\alpha = \prod_{r=1}^{|\alpha|} F_{w_r} \, , \qquad F_\Phi = \prod_{k=1}^S \left( F_{\alpha_k} \right)^{m_k} \, .
\label{eq:multiloop-annihilation}
\ee
Creation operators are obtained by taking hermitian conjugates. The multiloop creation operators $F_\Phi^\dagger$ are the basic ingredient for the construction of the loop expansion of physical states in loop quantum gravity which will be discussed in Section \ref{sec:loop-expansion}.

The seed space $\mathcal{H}_i$ carries a unitary representation of the group $SU(2)$, with generators $\vec{J}_i$ and Casimir operator $I_i$ defined by the quadratic expressions
\begin{equation}
\vec{J}_i\equiv\frac{1}{2}\vec{\sigma}_{AB}\,a_i^A{}^\dagger\, a_i^B,\qquad I_i\equiv \frac{1}{2}\delta_{AB}\;a_i^A{}^\dagger\, a_i^B\,.
\label{eq:J&I}
\end{equation}
Here $\vec{\sigma}_{AB}$ are Pauli matrices, and indices $A,B$ are raised, lowered and contracted always with the identity matrix $\delta_{AB}$. The generators satisfy the usual commutations relations:
\begin{equation}
[J_i^a,J_i^b] = i \epsilon^{ab}{}_c \, J_i^c \, .
\label{eq:J-commutators}
\end{equation}
The square of the $SU(2)$ generators is  $\vec{J}_i\cdot \vec{J}_i=I_i\,(I_i+1)$. We follow the standard notation and call \emph{spins} $ j_i=0,\frac{1}{2},1,\frac{3}{2},\ldots$ the eigenvalues of $I_i$. 

To each link $\ell=(s,t)$ of the bosonic lattice we associate a $2\times2$ operator matrix $h_\ell$ called the \emph{holonomy} and defined as:
\begin{equation}
(h_\ell)^A{}_B\equiv 
(2 I_t+1)^{-\frac{1}{2}}\big(
\epsilon^{AC}\, a_{t\,C}^\dagger\, a_{s\,B}^\dagger
-\epsilon_{BC}\, a_t^A\, a_s^C
\big)\,
(2 I_s +1)^{-\frac{1}{2}} \, .
\label{eq:hl}
\end{equation} 
Together with the $SU(2)$ generators $\vec{J}_i$, this operator satisfies the commutation relations
\begin{equation}
[\vec{J}_s,h_\ell]=\frac{1}{2}h_\ell\,\vec{\sigma},\qquad [\vec{J}_t,h_\ell]=-\frac{1}{2}\vec{\sigma} \, h_\ell\,.
\label{eq:Jh-commutators}
\end{equation}
Moreover, on the subspace of $\mathcal{H}_{\mathcal{S}}$ where the condition $I_{s(\ell)}=I_{t(\ell)}$ is satisfied, the holonomy operator commutes with itself:
\begin{equation}
[\,(h_\ell)^A{}_B,(h_{\ell '})^C{}_D]=0 \, .
\label{eq:h-commutators}
\end{equation}
Therefore, the operators $\vec{J_i}$ and $h_\ell$ introduced in Eqs.~\eqref{eq:J&I} and \eqref{eq:hl} correspond to a representation of the holonomy-flux algebra of observables of loop quantum gravity, defined by the Eqs.~\eqref{eq:J-commutators}, \eqref{eq:Jh-commutators} and \eqref{eq:h-commutators}, in the subspace of the bosonic Hilbert space $\mathcal{H}_{\mathcal{S}}$ selected by the condition $I_{s(\ell)}=I_{t(\ell)}$.

A bosonic representation of the holonomy-flux algebra has been first introduced in \cite{Livine:2013wmq}. The representation is not unique, however, and our formula for the holonomy operator differs from that presented in \cite{Livine:2013wmq}. The ambiguity is related to an arbitrary choice of factor ordering in the holonomy formula \eqref{eq:hl}: the commutation relations \eqref{eq:Jh-commutators} and \eqref{eq:h-commutators} are satisfied in the subspace with $I_{s(\ell)}=I_{t(\ell)}$ for any holonomy operator of the form:
\begin{equation}
\left[ h^{(\alpha)}_\ell \right]^A{}_B\equiv 
(2 I_t+1)^{\alpha}\big(
\epsilon^{AC}\, a_{t\,C}^\dagger\, a_{s\,B}^\dagger
-\epsilon_{BC}\, a_t^A\, a_s^C
\big)\,
(2 I_s +1)^{-1-\alpha}, \qquad \alpha \in \mathbb{R} \, .
\label{eq:order-ambiguity}
\end{equation}
The representation introduced in \cite{Livine:2013wmq} corresponds to $\alpha=0$, while our expression corresponds to the symmetric ordering $\alpha=-1/2$. A unique feature of the symmetrically ordered representation consists in that any eigenstate of the holonomy-operator corresponds to a delta function peaked at the associated eigenvalue $g \in SU(2)$ when mapped to the usual holonomy space representation of loop quantum gravity. More precisely, we have the following.

The Hilbert space of kinematical states of loop quantum gravity on a fixed graph $\Gamma$ in the holonomy representation is given by $\mathcal{K}_\Gamma = L^2[SU(2)]^{\otimes L}$, where a space $\mathcal{H}_\ell=L^2[SU(2)]$ of square integrable functions over $SU(2)$ is associated with each link $\ell$ in the graph $\Gamma$. An orthonormal basis of $\mathcal{H}_\ell$ is provided by the full set of normalized Wigner matrices $\sqrt{2j+1} \left[ D^j(g)\right]^m{}_n$. Holonomy operators act as multiplication operators:
\begin{equation}
(h_\ell)^A{}_B \, \psi(g_1,\dots,g_L) = (g_\ell)^A{}_B \, \psi(g_1,\dots,g_\ell) \, .
\label{eq:holonomy-rep}
\end{equation}
In the bosonic picture, on the other hand, a link $\ell$ is described by the Hilbert space of a system of four oscillators, $\mathcal{H}_{s(\ell)} \otimes \mathcal{H}_{t(\ell)}$, constrained by the condition $I_{s(\ell)}=I_{t(\ell)}$. This subspace is spanned by an orthonormal basis of states of the form:
\begin{equation}
\ket{j,m,n} = \frac{\bigl( a_t^{0 \dagger} \bigr)^{j+m} \bigl( a_t^{1 \dagger} \bigr)^{j-m} }{\sqrt{(j+m)! (j-m)!}} \frac{\bigl( a_s^{0 \dagger} \bigr)^{j+n} \bigl( a_s^{1 \dagger} \bigr)^{j-n} }{\sqrt{(j+n)! (j-n)!}} \ket{0} \, .
\label{eq:schwinger-basis}
\end{equation}
The unitary map between the holonomy and bosonic representations is defined by its action on the basis elements of the local spaces $\mathcal{H}_\ell$ \cite{Livine:2011gp,Livine:2013wmq}:
\begin{equation}
\sqrt{2j+1} \left[ D^j(g)\right]^m{}_n \mapsto (-1)^{j+n} \ket{j,m,-n} \, .
\label{eq:hol-bosonic-map}
\end{equation}
Under this map, the Dirac delta function peaked on an element $g \in SU(2)$ in the holonomy representation translates into
\begin{equation}
\ket{g}=\sum_j \frac{\sqrt{2j+1}}{(2j)!} (\epsilon_{AB} \bar{g}^A{}_C \, a_s^{C \dagger} a_t^{B \dagger})^{2j} \ket{0} \, ,
\label{eq:dirac-delta}
\end{equation}
and this can be easily checked to be an eigenstate of the symmetrically ordered holonomy operator:
\begin{equation}
(h_\ell)^A{}_B \ket{g} = g^A{}_B \ket{g} \, .
\label{eq:hl-eigenstates}
\end{equation}

The Wilson loop operator $W_\alpha$ associated with a closed loop $\alpha=\{\ell_1^{\epsilon_1},\ldots, \ell_{|\alpha|}^{\epsilon_{|\alpha|}}\}$ can now be constructed as usual by taking the trace of the product of link holonomies along the loop:
\be
W_\alpha = \mathrm{tr} \left( h_{|\alpha|} h_{|\alpha |-1} \cdots h_1 \right) \,,
\label{eq:wilson-loop}
\ee
where $h_i$ is the holonomy operator of the link $\ell_i^{\epsilon_i}$. More generally, we introduce operators associated with multiloops $\Phi=\{\alpha_1^{m_1} , \dots, \alpha_S^{m_S}\}$ as:
\be
W_\Phi = \prod_{k=1}^S W_{\alpha_k}^{m_k} \, .
\ee
Note that a normal ordering operator is naturally defined in the bosonic representation: the normal ordered operator $\normord{f(a^\dagger,a)}$ is obtained from $f(a^\dagger,a)$ by moving all annihilation operator to the right of all creation operators in the power series expansion of $f$, which we assume to exist. It turns out that the action of the normal ordered Wilson loop operators on the vacuum has the simple form
\be
\normord{W_\Phi} \ket{0} = F^\dagger_\Phi \ket{0} \, .
\ee
If $\Phi$ has any trivial tail of the form $\ell \cdot  \ell^{-1}$ in which a link is successively traversed back and forth, then $F_\Phi=0$. For a loop $\alpha$ without such tails and with at most one excitation per link, $F_\alpha^\dagger \ket{0}$ corresponds precisely to the Wilson loop state $\ket{\alpha}$ in the usual loop representation of loop quantum gravity \cite{Rovelli:2004tv,Gambini:1996ik}.

\subsection{Projection onto gauge-invariant space}

In order to determine the loop quantum gravity Hilbert space $\mathcal{H}_\Gamma$ associated with the graph $\Gamma$ we introduce two sets of constraints: 
\begin{equation}
C_\ell\equiv I_{s(\ell)}-I_{t(\ell)}\;\approx\;0\;,\qquad \vec{G}_n\equiv \sum_{i\in n}\vec{J}_i\;\approx\;0\,.
\label{eq:link-constraint}
\end{equation}
The \emph{link constraint} $C_\ell$ imposes the matching of the spins $j_s=j_t$ at the source and target of a link $\ell=(s,t)$. The \emph{node constraint} $\vec{G}_n$ imposes that the coupling of the $SU(2)$ representations associated with seeds at the node $n$ is invariant under overall $SU(2)$ transformations, i.e., the node is an intertwiner. These two sets of constraints can be implemented via projectors $P_\ell$ and $P_n$ so that the projector from the bosonic Hilbert space to the Hilbert space of loop quantum gravity is
\begin{equation}
P_\Gamma:\mathcal{H}_\mathcal{S}\to\mathcal{H}_\Gamma \, , \qquad\textrm{with}\quad \textstyle P_\Gamma=\Big(\prod_{n\in \Gamma} P_n \Big)\;\Big(\prod_{\ell \in \Gamma} P_\ell\Big)\,.
\label{eq:projection-prod}
\end{equation}
Note that the bosonic vacuum $|0\rangle$ is left invariant by the projector $P_\Gamma$ and therefore it belongs to the loop quantum gravity Hilbert space. This is the state with vanishing spins at all seeds, $\vec{J}_i\,|0\rangle=0$, $\forall i$, and therefore coincides with the \emph{Ashtekar-Lewandowski vacuum} on the graph $\Gamma$. 

Now let us discuss some explicit formulas for the projectors. The link projector $P_\ell$ can be obtained by averaging the link constraint $C_\ell$ over the group $U(1)$:
\be
P_\ell = \frac{1}{4\pi} \int_0^{4 \pi} d\phi \exp[- i \phi C_\ell] \, .
\ee
Computing this integral explicitly, we find an expression for the projector in the bosonic representation:
\begin{align}
P_\ell &= \normord{\mathcal{I}_0 \left( 4  \sqrt{I_s I_t} \, \right) \exp[-2(I_s + I_t) ] } 
\label{eq:cl-bosonic} \\
	 &= \normord{\exp[-2(I_s + I_t)] \sum_{k = 0}^\infty \frac{(4 I_s I_t)^k}{(k!)^2}} \; ,
\label{eq:cl-bosonic-series}
\end{align}
where $\mathcal{I}_\alpha$ denotes the modified Bessel functions of the first kind which have a series expansion:
\be
\mathcal{I}_\alpha(x) = \sum_{n=0}^\infty  \frac{1}{n! (n+\alpha)!} \left(\frac{x}{2}\right)^{2n + \alpha}.
\ee
Note that the square root appearing in \eqref{eq:cl-bosonic} is merely formal since all square roots drop out when the function is expanded in a series. Note also that the $k$-th term in the series \eqref{eq:cl-bosonic-series} projects onto spin $k/2$ for the link under consideration, allowing us to write:
\be
P_\ell = \sum_{j_\ell} P_{j_\ell} \, , \qquad P_{j_\ell}\equiv \normord{\exp[-2(I_s + I_t)]  \frac{(4 I_s I_t)^{2 j_\ell}}{(2 j_\ell!)^2}} \; .
\label{eq:Pjl-projector}
\ee

The projector $P_\ell$ can also be expressed in terms of its diagonal coherent state expansion.  For any family of complex numbers $z=\{z_i^A\}$, a coherent state $\ket{z} \in \mathcal{H}_{\mathcal{S}}$ in the bosonic representation is defined as usual by $a_i^A\ket{z}=z_i^A \ket{z}$. Normalized coherent states are given in explicit form as:
\be
\ket{z} = \exp\left(-\frac{1}{2} z_i^A z^i_A  \right) e^{z^i_A a_i^{A \dagger}} \ket{0} \, .
\ee
Restricting to a link $\ell=(s,t)$, a coherent state is then characterized by a multi-spinor $z_\ell=(z_s^A,z_t^B)$ with squared norm $|z_\ell|^2=|z_s|^2+|z_t|^2$. The link projector can be written as
\be 
P_\ell = \int d^4 z_\ell \, d^4 \bar{z}_\ell \; |z_\ell\rangle \langle z_\ell|\, e^{|z_\ell|^2} \, p_\ell(z_s,z_t) \, ,
\label{eq:cl-coherent-1}
\ee
with $p_\ell(z_s, z_t)$ given by
\begin{equation}
p_\ell(z_s, z_t) = \sum_{j=0}^\infty \frac{1}{(j!)^2} \left( \delta^{AB} \delta^{CD}\partial_{z^A_s}\partial_{\bar{z}^B_s}\partial_{z^C_t}\partial_{\bar{z}^D_t}\right)^j\, \delta (z_\ell,\bar{z}_\ell).
\label{eq:cl-coherent-2}
\end{equation}
Again, the $j$-th term in this series projects onto spin $j/2$.

For a given node, the projection operator $P_n$ can be obtained by group averaging over the $SU(2)$ gauge transformations generated by the exponentiation of the Gauss constraint $\vec{G}_n$,
\be
P_n = \int_{SU(2)} dg \, U_n(g) \, .
\label{eq:cn-bosonic-def}
\ee
In the bosonic representation, the action of a gauge transformation $g \in SU(2)$ at the node $n$ is given by the unitary transformation $U_n(g)$ such that
\be
U_n(g) a_i^{A \dagger} U^\dagger_n(g) = a_i^{\dagger B} g_B{}^A
\ee
for each seed $i=1,\dots,|n|$ at the node $n$, and which acts trivially in the remaining oscillators. The flux operators $J_i^a$ transform as:
\be
U_n^{\dagger}(g) J_i^a U_n(g) = [R(g)]^a{}_b J_i^b \, ,
\ee
where $[R(g)]^a{}_b \in SO(3)$ is the rotation determined by $g$. If $m,n$ are the nodes containing the source and target seeds of a link $\ell$, respectively, then the holonomy operator transforms as:
\be
[U_m^\dagger(g_s) U_n^\dagger(g_t)]  (h_\ell)^A{}_B [U_m(g_s) U_n(g_t)] = (g_t)^A{}_C (h_\ell)^C{}_D  (g_s^{-1})^D{}_B \, ,
\ee
as expected. Moreover, defining the node multi-spinor $z_n=(z_{n1},\dots,z_{n|n|})$ formed by all spinors at the node $n$, the action of a gauge transformation in a coherent state is simply $U_n(g)\ket{z}=\ket{g_n z}$, where $(g_n z)_i = g z_i$ for any seed $i=(n,\mu)$ at $n$, while spinors at the remaining nodes are not affected by the transformation.

Computing the integral \eqref{eq:cn-bosonic-def} explicitly, we obtain for the projection of the node constraint:
\begin{align} 
P_n &= \normord{ \frac{2 \mathcal{I}_1\Big(\sqrt{2F^{_\dagger}_{{ ij}} F^{{ij}}}\hspace{1pt}\Big)}{\sqrt{2F^{_\dagger}_{{ij}}F^{{ij}}}} \exp\left(-2 \sum_{i=1}^{|n|} I_i\right) } \nonumber \\
&= \normord{ \exp \left(- 2 \sum_{i = 1}^{|n|} I_i\right) \sum_{J=0}^\infty \frac{1}{J!(J+1)!}\left(\frac{F^\dagger_{ij} F^{ij}}{2}\right)^J } \, .
\label{eq:cn-bosonic}
\end{align}
The summations over $i$ and $j$ extend over all oscillators associated with the node of interest. Note that the node projector consists of a sum of orthogonal projectors onto the the gauge invariant subspaces with fixed total $J$ value:
\be
P_n = \sum_{J=0}^\infty P_J \, , \qquad P_J = \normord{ \frac{1}{J! (J+1)!}\left(\frac{F^\dagger_{ij} F^{ij}}{2}\right)^J  \exp\left(-2 \sum_{i=1}^{|n|} I_i \right) } \, .
\label{eq:PJn-projector}
\ee
The node projector also has a diagonal coherent state expansion:
\be
P_n = \frac{1}{\pi^{2|n|}} \int d^{2|n|}z_n \, d^{2|n|}\bar{z}_n \; |z_n\rangle \langle z_n|\, e^{|z_n|^2} p_n(z_n,\bar{z}_n) \, ,
\label{eq:cn-coherent-1}
\ee
where
\be
p_n(z_n,\bar{z}_n) = \sum_{J=0}^\infty \frac{1}{J!(J+1)!} \left( \sum_{i,j \in n} \frac{1}{2}\epsilon^{AB} \epsilon^{CD} \partial_{z^A_i} \partial_{z^B_j} \partial_{\bar{z}^C_i} \partial_{\bar{z}^D_j}\right)^J \delta(z_n,\bar{z}_n) \, .
\label{eq:cn-coherent-2}
\ee

\subsection{Spin network basis and bosonic representation}

The Hilbert space $\mathcal{H}_\Gamma$ of gauge-invariant states of loop quantum gravity on a fixed oriented graph $\Gamma$ admits an orthonormal basis labeled by spins $j_\ell$ and intertwiners $\mathrm{i}_n$, the \emph{spin-network} basis. A basis element $\ket{\Gamma,j_\ell,\mathrm{i}_n} \in \mathcal{H}_\Gamma$ in this representation is constructed as follows. A half-integer spin $j_\ell$ is first assigned to each link $\ell$ of the graph. Let $V^{j_\ell}$ be the corresponding irreducible representation of $SU(2)$ and $V^{j_\ell *}$ its dual representation. A representation space $V_{s(\ell)}=V^{j_\ell}$ is then attached to each source seed $s(\ell)$, and a dual representation $V_{t(\ell)}=V^{j_\ell *}$ to each target seed $t(\ell)$. Taking the tensor product of all such representations at a given node $n$, we obtain a reducible representation $V_n(\{j_\ell\})=\bigotimes_{\mu=1}^{|n|} V_{(n,\mu)}$ associated with the node, where the index $\mu$ labels links meeting at $n$ and the pair $i=(n,\mu)$ represents the corresponding seed. Spins are naturally assigned to the seeds according to $j_{s(\ell)}=j_{t(\ell)}=j_\ell$. An intertwiner $\mathrm{i}_n \in V_n(\{j_\ell\})$ is a state invariant under the action of $SU(2)$ on $V_n(\{j_\ell\})$. Expanding it in the standard magnetic number basis, we can write:
\be
\mathrm{i}_n = \sum_{m_{\mu}=-j_{(n,\mu)}}^{j_{(n,\mu)}}  \mathrm{i}_n^{m_1 \cdots m_a}{}_{m_{a+1} \cdots m_{|n|}} \, e^{j_{(n,1)}}_{m_1} \cdots e^{j_{(n,a)}}_{m_a}  \, e_{j_{(n,(a+1))}}^{m_{a+1}} \cdots e_{j_{(n,|n|)}}^{m_{|n|}}  \, ,
\ee
where $e_m^j$ is a basis element of the representation $V^j$, and $e^m_j$ a basis element of the dual representation. The number $a$ of upper indices in $\mathrm{i}_n$ corresponds to the number of links pointing outwards from the node; the lower indices correspond to links pointing towards the node.
 A spin network state is defined in the holonomy representation as:
\be
\scalar{g_\ell}{\Gamma,j_\ell,\mathrm{i}_n} = \sum_{m_{(n,\mu)}=-j_{(n,\mu)}}^{j_{(n,\mu)}} \left(\prod_n \mathrm{i}_n^{m_{(n,1)} \cdots}{}_{m_{(n,a_n+1)} \cdots } \right) \left( \prod_\ell \sqrt{2j_\ell+1} \left[D^{j_\ell}(g_\ell)\right]^{m_{t(\ell)}}{}_{ n_{s(\ell)}} \right) \, ,
\label{eq:spin-network-def}
\ee
Note that there is one contraction of indices for each seed $i$; however, a seed is represented as a pair $(n,\mu)$ when it appears as an intertwiner index, and as the target or source of a link, $t(\ell)$ or $s(\ell)$, when it appears as an index of a Wigner matrix. The contractions just follow the structure of the graph. The factor $\sqrt{2j_\ell+1}$ is a normalization constant for each Wigner matrix.

The spin network states defined in Eq.~\eqref{eq:spin-network-def} are gauge-invariant, $\ket{\Gamma,j_\ell,\mathrm{i}_n} \in \mathcal{H}_\Gamma$. Moreover, for a given spin distribution $j_\ell$, the space of intertwiners at each node is finite dimensional, allowing one to choose a finite complete set of orthonormal intertwiners $\mathrm{i}_n^{(\alpha)}$ for each $n$. The family of spin networks $\{ \ket{\Gamma,j_\ell,\mathrm{i}_n^{(\alpha)}} \}$ obtained by varying the spin configuration and orthonormal intertwiners over all possible configurations forms an orthonormal basis of $\mathcal{H}_\Gamma$.

Using the unitary map from the holonomy representation to the bosonic representation given by Eqs.~\eqref{eq:schwinger-basis} and \eqref{eq:hol-bosonic-map}, we can represent a spin network basis element in terms of creation operators $a_i^A{}^\dagger$ acting on the bosonic vacuum $|0\rangle$ by:
\begin{equation}
 |\Gamma,j_\ell,\mathrm{i}_n\rangle=\sum_{m_i=-j_i}^{+j_i}\!\!\Big(\prod_n [\mathrm{i}_n]_{m_{(n,1)}\cdots m_{(n,|n|)}}\Big)\Bigg(\prod_{i=1}^{2L} \frac{(a_i^{0\dagger})^{j_i-m_i}}{\sqrt{(j_i-m_i)!}}\frac{(a_i^{1\dagger})^{j_i+m_i}}{\sqrt{(j_i+m_i)!}} \Bigg) \; |0\rangle \, .
\label{eq:spin-network-bosonic}
\end{equation}
In this expression, the indices of the intertwiners are lowered using the isomorphism $\epsilon_j: V^j \to V^{j *} $ defined by $v_m = (-1)^{j-m} v^{-m}$. The tensor $\mathrm{i}_n$ with all indices lowered is an intertwiner in $\bigotimes_{\mu=1}^{|n|} V^{j_{(n,\mu)} *}$. Note that the inverse isomorphism can be used to raise the second index of the Wigner matrices, which are then mapped into the bosonic representation according to
\be
\sqrt{2j+1} \left[ D^j(g)\right]^{mn} \mapsto \ket{j,m,n} \, .
\ee
This gives an alternative presentation of the unitary map defined in Eq.~\eqref{eq:hol-bosonic-map}.

The resolution of the identity in the spin-network basis,
\begin{equation}
P_\Gamma=\sum_{j_\ell}\sum_{\mathrm{i}_n} |\Gamma,j_\ell,\mathrm{i}_n\rangle\langle\Gamma,j_\ell,\mathrm{i}_n| \, ,
\label{eq:Pspin-network}
\end{equation}
provides another expression for the projector from the bosonic Hilbert space $\mathcal{H}_{\mathcal{S}}$ to the loop quantum gravity Hilbert space $\mathcal{H}_\Gamma$.


\section{Loop expansion of the projector}
\label{sec:loop-expansion}

In this section we derive a loop expansion of the projector $P_\Gamma: \mathcal{H}_\mathcal{S} \to \mathcal{H}_\Gamma$. Let us first state the main result. A non-repeating loop $\alpha$ is a loop such that no cyclic permutation of its links exist that leaves the loop invariant. A non-repeating multiloop $\Phi=\{\alpha_1^{m_1}, \alpha_2^{m_2},\dots\}$ is a collection of non-repeating loops $\alpha_i$ with multiplicities $m_i$. In what follows, the multiloops $\Phi$ are non-repeating except when explicitly mentioned. For any multiloop $\Phi$, we can construct the corresponding multiloop state:
\be
\ket{\Phi} = F_\Phi^\dagger \ket{0} \, .
\ee
These states satisfy the link and node constraints and span the Hilbert space of loop quantum gravity. A resolution of the identity in $\mathcal{H}_\Gamma$ in terms of such overcomplete system is given by:
\begin{equation}
P_\Gamma=\sum_\Phi \frac{1}{\prod_\ell (2j_\ell)!\; \prod_n(J_n+1)!}\;F_\Phi^\dagger|0 \rangle\langle 0 |F_\Phi \, ,
\label{eq:Ploop}
\end{equation}
where $2j_\ell = 2j_\ell(\Phi)$ is the multiplicity of the link $\ell$ in the multiloop  $\Phi$ and $J_n=\sum_{i\in n}j_i$. The sum runs over all non-repeating multiloops $\Phi$, where the orientation of each loop $\alpha$ is kept fixed. The arbitrary choice of loop orientations does not affect the expansion. The \emph{loop expansion} \eqref{eq:Ploop} provides a representation of the projector from $\mathcal{H}_{\mathcal{S}}$ to $\mathcal{H}_\Gamma$. Applying this projector to an arbitrary state in $H_S$, we can represent its physical part as a linear superposition of multiloop states. In what follows, we first present a derivation of \eqref{eq:Ploop}, and then discuss alternative representations of the resolution of the identity in terms of combinatorial structures closely related to the multiloops $\Phi$, which we call symmetrized multiloops and routings.

\subsection{Multiloop representation}
\label{subsec:multiloop}

For each link $\ell$, the projector $\mathcal{P}_{j_\ell}$ to the space of states with spin $j_\ell$ in the space of solutions of the link constraint $C_\ell$ is given by Eq.~\eqref{eq:Pjl-projector}. Similarly, we defined in Eq.~\eqref{eq:PJn-projector} the projector $\mathcal{P}_{J_n}$ to the space of states with total spin $J_n$ in the space of solutions of the node constraint $\vec{G}_n$. Carrying combinatorial factors in the expressions \eqref{eq:Pjl-projector} and \eqref{eq:PJn-projector} to the left-hand side and then summing over the spins, we obtain the compact expressions:
\begin{align}
\sum_{J_n} \left[\prod_n(J_n+1)! \, \mathcal{P}_{J_n} \right] &= \;  \normord{ \exp\left(\sum_n \sum_{i,j \in n} \frac{1}{2} F_{ij}^\dagger F_{ij} - 2 \sum_i I_i\right)} \; , 
\label{formal_proj1}
\\
\sum_{j_\ell}\left[ \prod_\ell(2j_\ell)! \, \mathcal{P}_{j_\ell} \right] &= \; \normord{ \exp\left(2 I^{i}L_{ij} I^{j} - 2 \sum_i I_i\right)} \; ,
\label{formal_proj2}
\end{align}
where we have introduced the link connectivity matrix
\begin{equation}
L_{ij} = 
\begin{cases}
  1   & \text{if } i,j \text{ are connected by a link} , \\
  0   & \text{otherwise}.
 \end{cases}
\end{equation}
Multiplying the expressions (\ref{formal_proj1}) and (\ref{formal_proj2}) together and inserting a coherent state resolution of the identity between them we obtain:
\begin{align}
& \sum_{j_\ell} \left[\prod_n (J_n+1)! \right] \left[\prod_\ell (2j_\ell)!\right] \mathcal{P}_{\{j_\ell\}}\\
& \quad = \int \frac{d^{4L}z \, d^{4L}\bar{z}}{\pi^{4L}} \, \normord{ \exp\left( -z^A_i \bar{z}^i_A + \sum_n \sum_{i,j\in n} \frac{1}{2}z^i_A(\epsilon^{AB}F^\dagger_{ij})z_B^j  + \frac{1}{2}\bar{z}_A^i (a^A_i L_{ij} a^B_j)\bar{z}_B^j - a^{A\dagger}_i a^i_A\right) }
\label{eq:integral-projector}\\
&\quad = \normord{ \det(\mathbbm{1} - W L)^{-1/2} \exp( - a^{A\dagger}_i a^i_A) } \; ,
\end{align}
where $\mathcal{P}_{\{j_\ell\}}$ is the projector onto the space of physical states with spin configuration $\{j_\ell\}$, and we have defined a ``wedge'' matrix with operator entries
\begin{equation}
W_{ij} \equiv
\begin{cases}
  F^\dagger_{ij}F_{ij}  & \text{if } i,j \text{ belong to the same node}, \\
  0 & \text{otherwise}.
 \end{cases}
\end{equation}
We then rewrite the determinant in terms of the trace of a logarithm and expand the logarithm as a power series:
\begin{align}
\sum_{j_\ell} \left[\prod_n (J_n+1)! \right]\left[\prod_\ell (2j_\ell)!\right] \mathcal{P}_{\{j_\ell\}} & = 
 \normord{ \exp\left(\sum_{n=1}^\infty \frac{\text{tr}(WL)^n}{2n}- a^{A\dagger}_i a^i_A\right)} \nonumber \\
 &= \normord{ \exp\left(\sum_{\tilde{\alpha}} \frac{F^\dagger_{\tilde{\alpha}} F_{\tilde{\alpha}}}{R_{\tilde{\alpha}}}- a^{A\dagger}_i a^i_A\right)} \nonumber \\
&= \normord{ \exp\left(\sum_{\alpha} \sum_{R_\alpha=1}^\infty \frac{(F^\dagger_\alpha F_\alpha)^{R_\alpha}}{R_\alpha}- a^{A\dagger}_i a^i_A\right)} \nonumber \\
& = \normord{ \frac{\exp(-a^{A\dagger}_i a^i_A)}{\prod_\alpha (1 - F^\dagger_\alpha F_\alpha)} } \nonumber\\
& = \sum_{\{m_\alpha\}} \left[\prod_\alpha (F^\dagger_\alpha)^{m_\alpha}\right] |0\rangle \langle 0|\left[\prod_\alpha (F_\alpha)^{m_\alpha}\right] \nonumber \\
&= \sum_\Phi F^\dagger_\Phi |0\rangle \langle 0|F_\Phi. 
\label{eq:loop-expansion-der}
\end{align}
In the second line we have used the fact that $\text{tr}(WL)^n/2n$ generates all loops of length $n$, including loops that repeat, and divides by the number of repetitions $R_{\tilde{\alpha}}$. A tilde was added on the loops $\tilde{\alpha}$ to indicate that they are allowed to repeat. To obtain the third line, we have rewritten the sum in terms of non-repeating loops $\alpha$.  As a result, the multiloops appearing in the final result are composed of non-repeating loops.

We now project both sides of the Eq.~\eqref{eq:loop-expansion-der} at fixed $j_\ell$, carry the combinatorial factor from the left-hand side to right-hand side, and sum over $j_\ell$ to obtain
\begin{equation}
P_{\Gamma} = \sum_\Phi\frac{1}{\prod_n (J_n+1)! \prod_\ell (2j_\ell)!} F^\dagger_\Phi |0\rangle \langle 0|F_\Phi.
\label{eq:loop-expansion}
\end{equation}
We make the following comments concerning \eqref{eq:loop-expansion}:
\begin{itemize}
\item
As mentioned above, the multiloops $\Phi$ appearing in the sum are composed of non-repeating loops $\alpha$. Non-repeating means that the sequence of oriented links $\alpha = \{\ell_1^{\epsilon_1},\ldots, \ell_k^{\epsilon_k}\}$ has no nontrivial cyclic symmetries. Geometrically, this means that it is impossible to put $\alpha$ in the form $\alpha=\beta^k$, with $\beta \subset \alpha$ and $k > 1$. To count as a cyclic symmetry, the links \textit{and} their orientations must repeat. A loop is permitted to intersect itself arbitrarily many times. It may even retrace parts of the loop multiple times, as long as this does not result in a cyclic symmetry of the sequence of oriented links.
\item
The sum over $\Phi$ does not count different orientations multiple times. Rather, one first fixes an orientation for each loop $\alpha$ and then uses this orientation in every multiloop appearing in the sum. Note that while the sign of $F_\Phi$ is orientation dependent, the resolution of the identity includes both $F^\dagger_\Phi$ and $F_\Phi$ and is thus insensitive to this sign. 
\item
There can exist $\Phi_1 \ne \Phi_2$ such that $F^\dagger_{\Phi_1} |0\rangle = F^\dagger_{\Phi_2} |0\rangle$. If desired, this redundancy can be eliminated by summing over multiloops that have been symmetrized along the links, as we discuss later in this paper.
\end{itemize}

\subsection{Routing representation}

\begin{figure}[h!]
  \centering
    \includegraphics[width=.20\textwidth]{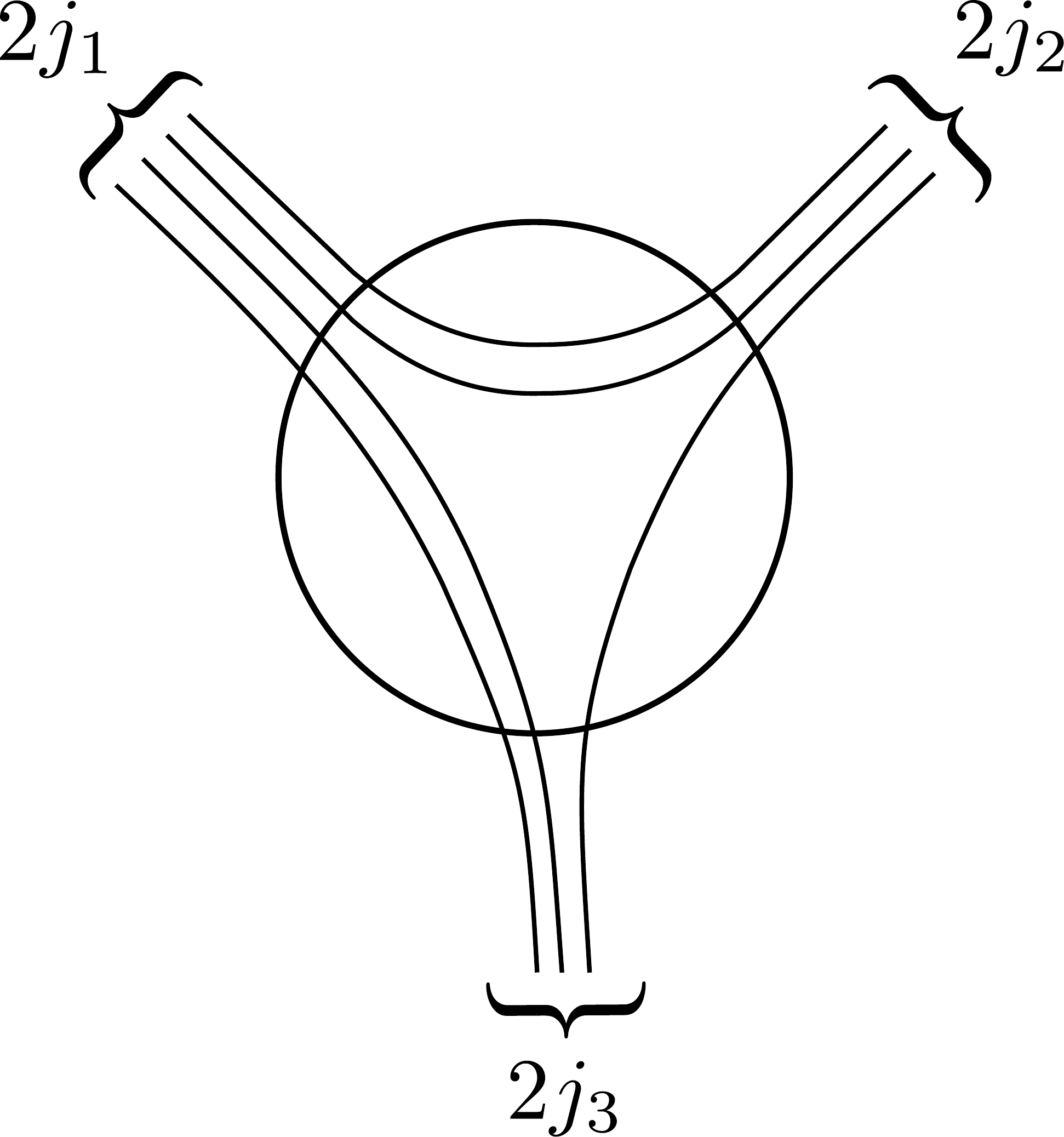}
     \caption{Links with spins $j_1 = 2, j_2 = 3/2, j_3 = 3/2$ decomposed into strands. The strands are connected at the node yielding a routing. The wedge multiplicities are: $n_{12}=2$, $n_{13}=2$, $n_{23}=1$.}
\label{fig:routing}
\end{figure}

Given a link with half-integer spin $j_\ell$, we can expand it into $2j_\ell$ \textit{strands}: $\{s_{\ell}^1,...,s_\ell^{2j_\ell}\}$. When links meet at a node, their strands can be connected in several ways. A complete pairing of all strands at a node is called a \textit{node routing} $R_n$. An example is shown in Fig.~\ref{fig:routing}.  A routing $ R=\{R_{n_1},R_{n_2},...\}$ is a full set of node routings, one for each node in $\Gamma$. We say that $R$ is a routing of $(\Gamma, j_\ell)$ when each link $\ell$ has exactly $2j_\ell$ strands in $R$, and write $R \in (\Gamma, j_\ell)$. 

A node routing $R_n$ determines a set $W_n(R)=\{{n_{ij}} \in \mathbb{N}; i,j \in n\}$, where the wedge multiplicity $n_{ij}$ counts the number of pairings of strands from the seeds $i,j \in n$ in the routing $R_n$. To each routing $R$, we assign a bosonic operator $F_R:\mathcal{H}_\mathcal{S} \to \mathcal{H}_\mathcal{S}$ defined by:
\be
F_R = \prod_n F_{R_n} \, , \qquad F_{R_n} = \prod_{i<j \in n} (F_{ij})^{n_{ij}} \, .
\label{eq:fr}
\ee
The operator $F_R$ is not uniquely determined by the routing $R$, since it also depends on the ordering of the seeds of the graph. Distinct orderings, however, can lead at most to a change of sign in $F_R$. Products involving an even number of occurrences of routing operators $F_R,F_R^\dagger$ are insensitive to this ambiguity and completely determined by the routing $R$.

We wish to prove that the projector $P_{\Gamma}:\mathcal{H}_\mathcal{S} \to \mathcal{H}_\Gamma$ admits the routing expansion:
\be
P_\Gamma=\sum_R \frac{1}{\prod_\ell [(2j_\ell)!]^2\; \prod_n(J_n+1)!}\;F_R^\dagger|0 \rangle\langle 0 |F_R \, .
\label{eq:routing-expansion}
\ee
We begin by focusing on a single node $n$. Let $H^{(n)}$ be the full bosonic Hilbert space at the node $n$, $H^{(n)}_J$ be the subspace of gauge-invariant states with fixed total area $J$, and $H^{(n)}_{j_\ell}$ be the subspace of gauge-invariant states with fixed spins $j_\ell$. Now define:
\begin{align}
\mathcal{P}_{(n,j_\ell)} &= \text{projector from $H^{(n)}$ to $H^{(n)}_{j_\ell}$} \, ,\\
\mathcal{P}_{(n,J)} &= \text{projector from $H^{(n)}$ to $H^{(n)}_{J}$} \, ,
\end{align}
and introduce the operators:
\begin{align}
\mathcal{O}_{(n,j_\ell)} &\equiv  \sum_{R_n \in (n,j_\ell)} \frac{1}{(J+1)! \prod_{\ell \in n} (2 j_\ell)!} F^\dagger_{R_n} |0 \rangle \langle 0 | F_{R_n} \, , 
\label{eq:O-j-ell}\\
\mathcal{O}_{(n,J)}  &\equiv \sum_{R_n \in (n,J)} \frac{1}{(J+1)! \prod_{\ell \in n} (2 j_\ell)!} F^\dagger_{R_n} |0 \rangle \langle 0 | F_{R_n} \, .
\label{eq:O-J-n}
\end{align}
The sum in Eq.~\eqref{eq:O-j-ell} runs over all routings with fixed spins $j_\ell$, while the sum in Eq.~\eqref{eq:O-J-n} runs over all routings with a fixed total spin $J$.
We will now prove that $\mathcal{O}_{(n,j_\ell)} = \mathcal{P}_{(n,j_\ell)}$ and $\mathcal{O}_{(n,J)} = \mathcal{P}_{(n,J)}$. In order to do this, it is enough to show that:
\begin{align*}
&1. \quad \mathcal{P}_{(n,J)} \ket{\psi} = 0 \implies \mathcal{O}_{(n,J)} \ket{\psi} = 0 , \quad \text{ for all } \ket{\psi} \in H^{(n)} \, , \\
&2. \quad \mathcal{O}_{(n,J)} \ket{\psi} \in H_J^{(n)} \, , \quad \text{ for all } \ket{\psi} \in H^{(n)} \, , \\
&3. \quad \bra{\phi} \mathcal{O}_{(n,J)} \ket{\psi} =  \bra{\phi} \mathcal{P}_{(n,J)} \ket{\psi}  ,  \quad \text{ for all } \ket{\phi},\ket{\psi} \in H^{(n)}_J \, .
\end{align*}
These three properties, when combined, imply that $\mathcal{O}_{(n,j_\ell)} = \mathcal{P}_{(n,j_\ell)}$.

Properties 1 and 2 follow trivially from the definition \eqref{eq:O-J-n}. Note that any state in $H^{(n)}$ can be decomposed as $\ket{\psi}=\mathcal{P}_{(n,J)} \ket{\psi} + (\mathbbm{1}- \mathcal{P}_{(n,J)}) \ket{\psi}$. If $\mathcal{P}_{(n,J)} \ket{\psi}=0$, then
\[
\bra{0} F_{R_n} \ket{\psi} = \bra{0} F_{R_n} (\mathbbm{1}- \mathcal{P}_{(n,J)}) \ket{\psi} =0 \qquad \text{ for } R_n \in (n,J) \, ,
\]
since $F^\dagger_{R_n} |0 \rangle \in H_J^{(n)}$, for all $R_n \in (n,J)$. But then $\mathcal{O}_{(n,J)} \ket{\psi} = 0$, from \eqref{eq:O-J-n}. The property 2 states that the image of $\mathcal{O}_{(n,J)}$ is a subspace of the image of $\mathcal{P}_{(n,J)}$, which follows immediately from \eqref{eq:O-J-n}.

It is sufficient to prove Property 3 for matrix elements between $U(N)$ intertwiners, since they form a complete set in $H^{(n)}_J$ \cite{Freidel:2010tt,Borja:2010rc}. Now, the wedge operators $F_{ij}$ have a simple action on $U(N)$ intertwiners $|J, \{z_i\} \rangle$,
\begin{equation}
F_{ij} \ket{J, \{z_i\}} = \sqrt{J(J+1)}z_{ij}|J-1, \{z_i\} \rangle \, ,
\end{equation}
where we have defined $z_{ij} \equiv \epsilon_{AB}z_i^A z_j^B$. Combining this with \eqref{eq:fr}, we find that:
\begin{equation}
\label{eqn}
F_{R_n}|J,\{z_i\}\rangle = \sqrt{J!(J+1)!} \prod_{ i<j} (z_{ij})^{n_{ij}} \ket{0} \, , \quad \text{for all } R_n \in (n,J)\, ,
\end{equation}
which yields:
\begin{align}
\langle J, \{w_i\} |\mathcal{O}_{(n,J)} \ket{J, \{z_i\} } &=   \sum_{R \in (n,J)} \frac{J!}{\prod_{\ell}(2j_\ell)!} \prod_{ i<j} (\bar{w_{ij}} z_{ij})^{n_{ij}} \nonumber \\
	&=  \sum_{\substack{n_{ij} \text{ with} \\ \sum n_{ij} = J}} \; \prod_{i<j} \frac{ J! }{n_{ij}!}( \bar{w_{ij}} z_{ij})^{n_{ij}} \nonumber \\
	&= \left(\sum_{i<j}  \bar{w}_{ij} z_{ij}\right)^J \nonumber \\
	&=  \langle J,\{w_i\}| J, \{z_i\} \rangle \, .
\label{eq:matrix-elements}
\end{align}
In the second line, we made the replacement $\sum_R  \to \sum_{n_{ij}}\prod_\ell (2j_\ell)!/\prod_{i<j} n_{ij}!$, where the extra factor counts the number of distinct routings $R$ that produce identical $n_{ij}$. In the last line we made use of known formulas for the inner product of the $U(N)$ intertwiners \cite{Freidel:2010tt,Borja:2010rc}. Since $U(N)$ intertwiners are gauge invariant, we have $\langle J,\{w_i\}| J, \{z_i\}\rangle  = \langle J,\{w_i\}|\mathcal{P}_{(n,J)}|J, \{z_i\}\rangle$, completing the proof of the Property $3$. We conclude that $\mathcal{O}_{(n,J)} = \mathcal{P}_{(n,J)}$.

Finally, for fixed spins $j_\ell$ such that the total spin at the node is equal to $J$:
\begin{align}
\mathcal{P}_{(n,j_\ell)} &= \mathcal{P}_{(n,j_\ell)}\mathcal{P}_{(n,J)} \mathcal{P}_{(n,j_\ell)}\\
 &=\mathcal{P}_{(n,j_\ell)}\mathcal{O}_{(n,J)} \mathcal{P}_{(n,j_\ell)}\\
 & =  \sum_{\substack{j_\ell' \text{ with }\\ \sum_{\ell} j_\ell' = J}} \mathcal{P}_{(n,j_\ell)}\mathcal{O}_{(n,j_\ell')}\mathcal{P}_{(n,j_\ell)}\\
& = \mathcal{O}_{(n,j_\ell)} \, .
\end{align}
We have thus shown that $\mathcal{O}_{(n,j_\ell)} = \mathcal{P}_{(n,j_\ell)}$. 

The projector $\mathcal{P}_{(\Gamma,j_\ell)}$ to the space of physical states with spin configuration $\{j_\ell\}$ in the full bosonic space is obtained by patching together multiple instances of $\mathcal{O}_{(n,j_\ell)}$, one for each node. Summing over all spin configurations, we obtain the projector $\mathcal{P}_\Gamma$ in the desired form \eqref{eq:routing-expansion}\footnote{Note the extra factor of $\prod_\ell (2j_\ell)!$ appearing in the denominator of \eqref{eq:routing-expansion} compared to \eqref{eq:O-j-ell}. Each link is counted twice since it belongs to two nodes.}.

Some remarks are now in order concerning the routing expansion \eqref{eq:routing-expansion} of the projector $\mathcal{P}_\Gamma$ and about its relation to the multiloop expansion \eqref{eq:loop-expansion} presented in the last section.

Similarly to what happens in the multiloop representation, in the sum over routings $R$ one does not count distinct orientations multiple times. Combinatorially, a routing $R$ is a list of non-oriented wedges connecting pairs of strands at each node. By assembling the strands and wedges together, we produce a series of loops in $\Gamma$, but these are not oriented. Notice that a routing $R$ cannot be directly identified with the multiloop $\tilde{\Phi}(R)$ it generates, since it also includes extra information on how the distinct strands at a link are crossed by the loops.

The multiloop $\tilde{\Phi}(R)$ in general contains repeating loops. A non-repeating multiloop $\Phi(R)$ is obtained by the simple procedure of breaking the repeating loops in $\tilde{\Phi}(R)$ into their elementary non-repeating pieces. For example, a loop $\alpha \cdot \alpha \in \tilde{\Phi}(R)$ formed by circling twice a non-repeating loop $\alpha$ corresponds to two occurrences of $\alpha$ in $\Phi(R)$. It turns out that
\be
F_{\Phi(R)}^\dagger|0 \rangle\langle 0 |F_{\Phi(R)} = F_R^\dagger|0 \rangle\langle 0 |F_R \, ,
\ee
allowing contributions from all routings $R$ of a multiloop $\Phi(R)$ to be grouped together in \eqref{eq:routing-expansion}. But we can prove that the number of routings producing a given multiloop $\Phi$ is given by $\prod_{j_\ell} (2j_\ell)!$ (see Appendix \ref{sec:count-routings}). Making then the replacement $\sum_R \to \sum_\Phi \prod_\ell (2j_\ell)(\Phi)$ in \eqref{eq:routing-expansion}, we recover the multiloop expansion \eqref{eq:loop-expansion}. This constitutes an independent proof of \eqref{eq:loop-expansion}. An advantage of this new proof of the multiloop expansion of the projector is that it involves only well-defined operators at all steps of the demonstration, while the arguments used in Section \ref{subsec:multiloop} involve formal manipulations of divergent operators. In this way, the shorter formal demonstration of the multiloop expansion previously discussed is here complemented by the combinatorially more involved proof based on the formalism of routings.

\subsection{Symmetrized multiloop representation}

The multiloops $\Phi$ that label individual terms in the loop expansion \eqref{eq:loop-expansion} of the projector $\mathcal{P}_\Gamma$ to the space of physical states are redundant. There exist many distinct $\Phi_1, \Phi_2$ for which $F_{\Phi_1} |0\rangle = (\pm) F_{\Phi_2} |0\rangle$. We can eliminate this redundancy by grouping the multiloops into equivalence classes. Given some multiloop $\Psi$, we say that $\Phi \sim \Psi$ if $\Phi$ can be obtained from $\Psi$ by permuting the loops as they pass through the links (i.e., by `cutting' along the links, and `rewiring' the loops together). A symmetrized multiloop $[\Psi]$ is an equivalence class of such multiloops:
\begin{equation}
[\Psi] = \{ \Phi | \Phi \sim \Psi \} \, ,
\label{eq:equiv-symm}
\end{equation}
and is fully specified by the multiplicities $n_{ij}$ of the wedges for any representative $\Psi$ of $[\Psi]$. We assign a bosonic operator to each $[\Psi]$ through:
\begin{equation}
F_{[\Psi]} = \prod_n \, \prod_{i<j \in n} (F_{ij})^{n_{ij}} \, .
\label{eq:F-psi-def}
\end{equation}
Note that for all $\Phi \in [\Psi]$, we have $F_\Phi = (\pm) F_{[\Psi]}$. The symmetrization of  loops along the links is thus naturally built into the loop expansion in the bosonic representation in the sense that, when one constructs multiloops states $F_\Phi^\dagger\ket{0}$, they automatically come out symmetrized.

In terms of symmetrized multiloops, the resolution of the identity becomes:
\begin{equation}
\mathcal{P}_{\Gamma}=\sum_{[\Psi]} \frac{1}{\prod_n(J_n+1)! \prod_{i<j \in n} n_{ij}!} F^\dagger_{[\Psi]}\ket{0} \bra{0} F_{[\Psi]} \, .
\label{eq:symmetrized-expansion}
\end{equation}
This can be obtained directly from \eqref{eq:routing-expansion} by noting that there are
\begin{equation}
\frac{\prod_\ell[(2j_\ell)!]^2}{\prod_{i<j \in n}n_{ij}!}
\end{equation}
routings corresponding to a given $[\Psi]$ (i.e., a given specification of wedges). The expression \eqref{eq:symmetrized-expansion} can also be obtained from Eq.~\eqref{eq:integral-projector} by directly computing the integral and collecting all terms associated with $[\Psi]$. The operator $F_{[\Psi]}$ depends not only on $[\Psi]$ but also on the labeling of the seeds, which determines its sign. The product $F^\dagger_{[\Psi]}\ket{0} \bra{0} F_{[\Psi]}$ is insensitive to this ambiguity, however, and the expansion \eqref{eq:symmetrized-expansion} is independent of the chosen labelling of the seeds.

The expansion \eqref{eq:symmetrized-expansion} allows us to write any physical state $\ket{\psi} \in \mathcal{H}_\Gamma$ as a superposition of symmetrized multiloop excitations $F_{[\Psi]}^\dagger \ket{0}$. Such a representation of the space of physical states closely resembles that used  for the introduction of the spin network basis in the original work \cite{Rovelli:1995ac}. Indeed, if we extend the equivalence relation \eqref{eq:equiv-symm} to include generic (repeating) multiloops and restrict to the case of trivalent graphs $\Gamma$, the agreement is complete. A spin network state is then labeled by a symmetrized multiloop $[\Psi]$ and corresponds to the bosonic excitation $F_{[\Psi]}^\dagger \ket{0}$ in $\mathcal{H}_\mathcal{S}$, up to an overall sign. If $\Gamma$ has nodes $n$ with valence larger than $3$, then we also need to choose an orthonormal basis of intertwiners $\mathrm{i}_n$ at each node to describe spin network states. An intertwiner corresponds to a particular linear superposition of configurations of wedges at each node. Attaching the wedges to the symmetrized strands at the links, we can represent $\ket{\Gamma,j_\ell,\mathrm{i}_n}$ as a superposition of symmetrized loop excitations in $\mathcal{H}_\mathcal{S}$. In short, every multiloop state is equivalent to some assignment of intertwiners to nodes, but not every assignment of intertwiners to nodes is a multiloop state.

In the representation provided by the resolution of the identity \eqref{eq:symmetrized-expansion}, in contrast, the overcompleteness of the loop basis is only partially solved. Let us recall that Wilson loop states satisfy the Mandelstam and retracing identities \cite{Rovelli:1995ac}:
\begin{align}
\ket{\alpha \cup \beta} &= \ket{\alpha \cdot \beta} + \ket{\alpha \cdot \beta^{-1}} \, , 
\label{eq:mandelstam} \\
\ket{\alpha} &= \ket{\alpha \cdot \gamma \cdot \gamma^{-1}} \, .
\label{eq:retracing}
\end{align}
The Mandelstam identity \eqref{eq:mandelstam} is valid for loops with a common basepoint, but extends to arbitrary pairs of loops when combined with the retracing identity \eqref{eq:retracing}:
\be
\ket{\alpha \cup \beta} = \ket{\alpha \cdot \gamma \cdot \beta \cdot \gamma^{-1}} + \ket{\alpha \cdot \gamma \cdot \beta^{-1} \cdot \gamma^{-1} } \, .
\label{eq:extended-mandelstam}
\ee
Consider now the bosonic excitations constructed with the $F^\dagger_\Phi$ operators. 
The first marked difference is that the retracing identity ceases to be valid. In fact, for any loop $\alpha'$ with a tail, $\alpha'= \alpha \cdot \gamma \cdot \gamma^{-1}$, we have $F^\dagger_{\alpha'}\ket{0}=0$. This means that the retracing identity is automatically solved in the bosonic formalism: only multiloops with all tails removed are included in the expansion \eqref{eq:symmetrized-expansion}\footnote{This can be seen explicitly from the restriction to $i<j$ in the product within each node in Eq.~\eqref{eq:symmetrized-expansion}. The presence of a tail is indicated by some nonzero multiplicity with $i=j$.}. As a result, Mandelstam identities of the form \eqref{eq:extended-mandelstam} are not present, and it suffices to consider the case of loops with a common basepoint. 

In particular, consider two loops which share a link $\gamma$, say $\alpha=\gamma \cdot \alpha_1$ and $\beta=\gamma \cdot \beta_1$. Then we have:
\be
\ket{\gamma \cdot \alpha_1 \cup \gamma \cdot \beta_1} = \ket{\gamma \cdot \alpha_1 \cdot \gamma \cdot \beta_1} + \ket{\gamma^{-1} \cdot \gamma \cdot \alpha_1 \cdot \beta_1^{-1}} \, ,
\ee
where we applied a cyclic translation to the second term on the right-hand side. But since this term has a tail, the corresponding $F$ operator is zero. Accordingly, we have in the  bosonic representation:
\be
F^\dagger_{\alpha \cup \beta} = \pm F^\dagger_{\alpha \cdot \beta} \, .
\ee
That is, the Mandelstam identity is automatically solved for loops which share a link, up to the sign convention. We are thus left with Mandelstam identities for loops which intersect at isolated nodes. These are encoded in the Pl\"ucker identities \cite{Bonzom:2012bn}
\be
F_{ij} F_{kl} = F_{ik} F_{jl} + F_{il} F_{kj}
\label{eq:plucker}
\ee
relating wedge operators at a node. Such identities describe the residual overcompleteness present in the basis of symmetrized loop excitations $F^\dagger_{[\Psi]} \ket{0}$. A given $[\Psi]$ is fully specified by a set of wedge multiplicities $n_{ij}$ at its nodes, but states associated with distinct $n_{ij}$'s are related by identities \eqref{eq:plucker}.

As seen from the wealth of applications of the usual coherent state representation in many-body problems, the overcompleteness of a basis is not an issue as far as a resolution of the identity is available. The loop expansions of the projector to the space of physical states discussed in this section provide just such a tool in the bosonic representation of loop quantum gravity. Instead of solving all Mandelstam and retracing identities, these are first reduced by construction to a smaller set associated with nodes of the graph. The resulting local notion of overcompleteness is then dealt with by the introduction of a resolution of the identity at each node. When extended to the full graph, this technique naturally leads to the loop expansions discussed in this section.


\section{Coherent and squeezed states}
\label{sec:semiclassical}

One of the most important applications of the bosonic representation lies in the definition of semiclassical states in loop quantum gravity. Since the construction of states with prescribed average values and correlation functions is straightforward for a system of harmonic oscillators, the construction of semiclassical states in $\mathcal{H}_\mathcal{S}$ poses no difficulties. Such states can be projected to the space of states of loop quantum gravity $\mathcal{H}_\Gamma \subset \mathcal{H}_\mathcal{S}$, leading in particular to the definition of coherent \cite{Dupuis:2010iq,Bonzom:2012bn} and squeezed \cite{Bianchi:2016} spin network states. The projection can be implemented using the loop expansion of the projector obtained in the last section, Eq.~\eqref{eq:Ploop}. This yields concrete representations of the projected states in the loop basis:
\begin{align}
\ket{\psi} 	&= P_\Gamma \ket{\psi_\mathcal{S}} \notag \\
		&= \sum_\Phi \frac{ \scalar{\Phi}{\psi_\mathcal{S}}}{\prod_\ell (2j_\ell)!\; \prod_n(J_n+1)!}\;F_\Phi^\dagger|0 \rangle \, ,
\end{align}
where $\ket{\psi_\mathcal{S}} \in \mathcal{H}_\mathcal{S}$ is some state in the bosonic representation. If the scalar products $\scalar{\Phi}{\psi_\mathcal{S}}$ can be computed for all multiloops $\Phi$, then the loop expansion of the state is completely determined. In this section we apply this technique to a variety of states, including the cases of coherent, squeezed and heat kernel states, for which explicit loop expansions will be written. We first briefly review the definition of such states.

\subsection{Coherent spin networks}
\label{sec:coherent}

In this section we provide the definitions in the bosonic representation of the most commonly encountered types of coherent states and summarize their inter-relations \cite{Livine:2007vk,Freidel:2010tt,Bonzom:2012bn,Borja:2010rc,Dupuis:2010iq,Bianchi:2009ky,Bianchi:2010mw}. A review of their properties can be found in \cite{Bonzom:2012bn}, to which we refer for details. Then we discuss the loop representation of coherent spin networks using the techniques introduced in Section \ref{sec:loop-expansion}.

A  spin coherent state \cite{Perelomov:1986tf,Gazeau:2009zz} in the irreducible representation $V^{j_i}$ of $SU(2)$ spanned by bosonic states with $2j_i$ excitations in the Hilbert space of the seed $i$ is characterized by a spinor $z_i^A \in \mathbb{C}^2$:
\be
\ket{j_i,z_i} = \frac{\bigl( z_A^i a_i^{A \dagger} \bigr)^{2j_i}}{\sqrt{(2 j_i)!}} \ket{0} \, .
\label{eq:su2-cs}
\ee
Its norm is given by $\scalar{j_i,z_i}{j_i,z_i}=(\delta_{AB} \bar{z}^A z^B)^{2j_i}$. Such states are peaked at $\mean{\vec{J}_i} = j_i \vec{v}(z_i)/|\vec{v}(z_i)|$,\footnote{The brackets represent the average $\mean{\vec{J}_i} = \bra{j_i,z_i} \vec{J}_i \ket{j_i,z_i}/\scalar{j_i,z_i}{j_i,z_i}$.} where the three-dimensional vector $\vec{v}(z)$ associated with a spinor $z$ is
\be
\vec{v}(z) = \frac{1}{2} \vec{\sigma}_{AB} \bar{z}^A z^B \, .
\ee
The $3$d vector $\vec{v}(z)$ has norm $|\vec{v}|=\sqrt{\vec{v}\cdot \vec{v}} = (1/2) \delta_{AB} \bar{z}^A z^B$. The spinor $z$ is determined by $\vec{v}(z)$ only up to a phase $e^{i \xi}$. Therefore, the spin coherent state $\ket{j_i,z_i}$ is determined, up to a phase and a normalization, by a direction $\hat{v}$ in the two-sphere $S^2$ and the value of the Casimir operator $\vec{J}\cdot \vec{J}=j_i(j_i+1)$. It provides the semiclassical description of a state with angular momentum $j_i \hat{v}$.

A Livine-Speziale (LS) coherent state in $\mathcal{H}_\Gamma$ \cite{Livine:2007vk,Freidel:2010tt,Dupuis:2010iq} is characterized by a set of parameters  $\{j_i,z_i\}$ attached to the seeds $i$ of the graph $\Gamma$, where the $j_i$'s are spins and $z_i \in \mathbb{C}^2$. It is defined as the tensor product of local $SU(2)$ coherent states at seeds projected to the space of physical states:
\be
\ket{\{j_i,z_i\}} = P_\Gamma \bigotimes_{i=1}^{2L} \ket{j_i,z_i} \, .
\ee
Scalar products and norms of LS states are discussed in \cite{Bonzom:2012bn}. Note that these states have definite spins $j_\ell$ at the links of the graph, i.e., they are eigenstates of all spin operators $J_i$. As a result, the variables conjugate to the $J_i$'s are completely uncertain. Accordingly, LS states represent semiclassical states for the spatial (intrinsic) geometry, but are not peaked at any particular classical configuration of the extrinsic geometry \cite{Rovelli:2004tv}. 
 
A $U(N)$ coherent intertwiner $\ket{J_n,\{z_i\}}_n$ at a node $n$ is characterized by a non-negative integer $J_n \in \mathbb{N}_0$ and a set of spinors $z_i^A \in \mathbb{C}^2$ attached to the seeds $i=(n,\mu)$ at the node:
\be
\ket{J_n,\{z_i\}}_n = \frac{1}{\sqrt{J_n! (J_n+1)!}} \left( \frac{1}{2} \sum_{i,j \in n} \epsilon_{AB} z_i^A z_j^B F_{ij}^\dagger \right)^{J_n} \ket{0} \, .
\ee
It consists of a superposition of all LS states with total spin $J_n$ at the node:
\be
\ket{J_n,\{z_i\}}_n = \sqrt{J_n! (J_n+1)!} \sum_{\sum j_i = J_n} \frac{1}{\prod_{i \in n} \sqrt{(2j_i)!}} \ket{\{j_i,z_i\}} \, .
\ee
When the vectors $\vec{v}(z_i)$ satisfy the closure condition $\sum \vec{v}(z_i) = 0$, the $U(N)$ coherent intertwiner $\ket{J_n,\{z_i\}}_n$ provides a semiclassical picture of the convex polyhedron with $|n|$ faces specified by the unit normals $\hat{v}(z_i)=\vec{v}(z_i)/|\vec{v}(z_i)|$ and the total area $J_n$. This polyhedron describes the semiclassical geometry of the node $n$ in the dual lattice $\Gamma^*$. Since the spinors $z_i$ have more information than the normal vectors $\vec{v}(z_i)$, one ends up with an extra phase $e^{i \xi_i}$ attached to each link, resulting in a framed polyhedron \cite{Freidel:2010tt,Bianchi:2010gc}.

A coherent intertwiner $\ket{\{z_i\}}_n$ at the node $n$ is defined as the gauge-invariant projection of coherent states associated with the annihilation operators $a_i^A$ in the bosonic representation $\mathcal{H}_\mathcal{S}$:
\be
\ket{\{z_i\}}_n = P_n \bigotimes_{i \in n} e^{z_A^i a_i^{A \dagger}} \ket{0} \, .
\ee
It corresponds to a simple superposition of  $U(N)$ coherent intertwiners:
\be
\ket{\{z_i\}}_n = \sum_{J_n} \frac{1}{\sqrt{J_n! (J_n+1)!}} \ket{J_n,\{z_i\}}_n \, .
\ee
The construction naturally extends to the full graph $\Gamma$. A coherent spin network $\ket{\{z_i\}}$ is defined as the projection to the space of physical states of coherent states for the full set of harmonic oscillators in the bosonic representation $\mathcal{H}_\mathcal{S}$:
\be
\ket{\{z_i\}} = P_\Gamma \bigotimes_{i=1}^{2L} e^{z_A^i a_i^{A \dagger}} \ket{0} \, .
\label{eq:coherent-sn}
\ee

To compute the loop expansion of a coherent spin network $\ket{\{w_i\}}$, we first introduce a set of complex variables $z_i^A \in \mathbb{C}^2$ associated with the seeds $i$ of the graph $\Gamma$ and define the holomorphic function:
\be
Z_\Phi\equiv
\prod_{\alpha\in\Phi}\Big(\prod_{\langle i,j\rangle\in \alpha}\!\!\epsilon_{AB}\,z^A_i z^B_j\;\Big)^{m_\alpha}\,.
\label{eq:Zphi}
\ee
The scalar product of a coherent state $\ket{\{w_i\}}$ with a multiloop state $F_\Phi^\dagger|0\rangle$ defines a function $\rho_\Phi(w)$ that can be expressed as a complex integral
\begin{equation}
\rho_\Phi(w)\equiv\langle 0|F_\Phi \ket{\{w_i\}}=
\int \frac{d^{4L}z \, d^{4L}\bar{z}}{\pi^{4L}} \, Z_\Phi\;
e^{-z_i^A \bar{z}^i_A+\bar{z}_A^i w_i^A}  = Z_\Phi |_{z=w}\, .
\label{eq:rho-Phi}
\end{equation}
Using now the representation \eqref{eq:loop-expansion} of the projector $P_\Gamma$ in \eqref{eq:coherent-sn} and inserting a coherent state resolution of the identity in the resulting expression, we find that a coherent state $\ket{\{z_i\}}$ has the following loop expansion:
\begin{equation}
\ket{\{z_i\}} = \sum_\Phi \frac{Z_\Phi}{\prod_\ell (2j_\ell)!\; \prod_n(J_n+1)!}\;F_\Phi^\dagger|0\rangle.
\label{eq:loop-expansion-coherent}
\end{equation}

\subsection{Squeezed spin networks}
\label{sec:squeezed}

Squeezed vacua in loop quantum gravity have been recently introduced in \cite{Bianchi:2016}. They are defined as the projection to the space of physical states $\mathcal{H}_\Gamma$ of the usual squeezed vacua of bosonic systems. Following \cite{Bianchi:2015fra}, a squeezed vacuum state $\ket{\gamma}$ in the full bosonic Hilbert space $\mathcal{H}_\mathcal{S}$ is here labeled by a complex matrix $\gamma$ in the Siegel unit disk $\mathcal{D}$ defined as
\begin{equation}
\mathcal{D}=\{\gamma\in \textrm{Mat}(4L,\mathbb{C})|\,\gamma=\gamma^t\;\;\textrm{and}\;\; \mathds{1}-\gamma\gamma^\dagger>0\,\}\,.
\label{eq:Siegel}
\end{equation}
The squeezing matrix $\gamma$ uniquely determines an element $M_\gamma$ of the symplectic group $Sp(4L,\mathbb{R})$. The bosonic space $\mathcal{H}_\mathcal{S}$ carries a unitary representation of $Sp(4L,\mathbb{R})$, and the bosonic operators transform under $M_\gamma$ as:
\be
U(M_\gamma) \, a_i^A \, U(M_\gamma)^{-1} = \Phi^{ij}_{AB} \, a_j^B + \Psi^{ij}_{AB} \, a_j^{B \dagger} \, ,
\ee
with $\Phi = (1-\gamma \gamma^\dagger)^{-1/2}$ and $\Psi = (1-\gamma \gamma^\dagger)^{-1/2} \gamma$ \cite{Bianchi:2015fra}.  The squeezed vacuum $\ket{\gamma}$ is the result of the action of $U(M_\gamma)$ on the vacuum state in $\mathcal{H}_\mathcal{S}$:
\begin{align}
\ket{\gamma} 	& = U(M_\gamma) \ket{0} \nonumber \\
			&=  \det(\mathbbm{1}-\gamma \gamma^\dagger)^{1/4} \exp\Big(\frac{1}{2}\, \gamma_{AB}^{ij}\, F^{AB}_{ij}{}^\dagger\Big)\,|0\rangle\,.
\label{eq:squeezed-def}
\end{align}
In general, such states do not solve the link and node constraints \eqref{eq:link-constraint}. Squeezed vacua in loop quantum gravity are obtained by projecting them to the space of physical states $\mathcal{H}_\Gamma$:
\be
\ket{\Gamma,\gamma} = P_\Gamma \ket{\gamma} \in \mathcal{H}_\Gamma \, .
\label{eq:squeezed-lqg}
\ee
In what follows we shall omit the normalization factor $\det(\mathbbm{1}-\gamma \gamma^\dagger)^{1/4}$ in the definition of $\ket{\Gamma,\gamma}$, since the projection changes the norm of the state.

The loop expansion of a squeezed vacuum $\ket{\Gamma,\gamma}$ is obtained following the same procedure as for coherent states. The scalar product of a squeezed vacuum $\ket{\Gamma, \gamma}$ with a multiloop state $F_\Phi^\dagger|0\rangle$ defines a function $\mu_\Phi(\gamma)$ that can be expressed as a complex integral,
\begin{equation}
\mu_\Phi(\gamma)\equiv\langle 0|F_\Phi \ket{\Gamma,\gamma}=
\int \frac{d^{4L}z \, d^{4L}\bar{z}}{\pi^{4L}} \, Z_\Phi\;
e^{-z_i^A \bar{z}^i_A+\frac{1}{2} \gamma_{ij}^{AB} \bar{z}_A^i \bar{z}_B^j}  \, .
\label{eq:mu-Phi}
\end{equation}
Using the representation \eqref{eq:loop-expansion} of the projector $P_\Gamma$ in \eqref{eq:squeezed-lqg} and inserting a coherent state resolution of the identity in the resulting expression, we find the representation of $\ket{\Gamma,\gamma}$ as a superposition of multiloop excitations:
\begin{equation}
\ket{\Gamma,\gamma} = \sum_\Phi \frac{\mu_\Phi(\gamma)}{\prod_\ell (2j_\ell)!\; \prod_n(J_n+1)!}\;F_\Phi^\dagger|0\rangle.
\label{eq:loop-expansion-squezed}
\end{equation}
The coefficients $\mu_\Phi(\gamma)$ cannot be computed in closed form for arbitrary squeezing matrices $\gamma$, but that can be done for special classes of locally squeezed states and perturbations thereof as we will see later in the paper.

The projection $P_\Gamma$ in Eq.~\eqref{eq:squeezed-lqg} can be alternatively implemented using the resolution of the identity in the spin network basis (\ref{eq:Pspin-network}). This can be done by first introducing the holomorphic function
\begin{equation}
Z_{j_\ell,\,\mathrm{i}_n}=\sum_{m_i=-j_i}^{+j_i}\!\!\!\Big(\prod_n \left[ \bar{\mathrm{i}}_n \right]_{m_1\cdots m_{|n|}}\Big)
\Big(\prod_{i=1}^{2L}
{\textstyle
\frac{(z_i^0)^{j_i-m_i}}{\sqrt{(j_i-m_i)!}}\frac{(z_i^1)^{j_i+m_i}}{\sqrt{(j_i+m_i)!}}
}
\Big) \, ,
\end{equation}
where the $z_i^A$ are again complex variables attached to the seeds of the graph $\Gamma$. Next we define the $\gamma$-transform of this function as
\begin{equation}
 c_{\,\mathrm{i}_n,j_\ell}(\gamma)= \scalar{\Gamma,j_\ell,\mathrm{i}_n}{\Gamma,\gamma} =
\int \frac{d^{4L}z \, d^{4L}\bar{z}}{\pi^{4L}} \,Z_{j_\ell,\,\mathrm{i}_n}\;e^{-z_i^A \bar{z}^i_A+\frac{1}{2}\gamma_{ij}^{AB}\bar{z}^i_A \bar{z}^j_B} \,.
\end{equation}
The spin-network expansion of the squeezed vacuum $|\Gamma,\gamma\rangle$ is then given by the linear superposition
\begin{equation}
|\Gamma,\gamma\rangle=\sum_{j_\ell,\,\mathrm{i}_n}c_{j_\ell,\,\mathrm{i}_n}(\gamma)\;|\Gamma,j_\ell,\mathrm{i}_n\rangle.
\label{eq:squeezed-sn-basis}
\end{equation}

Yet another representation is obtained by writing the projection $P_\Gamma$ as the product of individual link and node projections, as in Eq.~\eqref{eq:projection-prod}, and then using the diagonal coherent representations \eqref{eq:cl-coherent-1} and \eqref{eq:cn-coherent-1} for $P_n$ and $P_\ell$. Integrating the expression so obtained, we find:
\be
\ket{\Gamma,\gamma} = \sum_{j_\ell} \frac{1}{\prod_n(J_n+1)!} D_{\{j_\ell\}} \exp\left( \frac{1}{2}  \gamma_{ij}^{AB} \bar{w}_A^i \bar{w}_B^j + \frac{1}{2} F_{ij}^\dagger \epsilon^{AB} w^i_A w^j_B\right) \ket{0} \Big|_{w=0} \, ,
\label{eq:squeezed-der-repr}
\ee
where
\be
D_{\{j_\ell\}} = \prod_{\ell=1}^L \frac{1}{[(2j_\ell)!]^2} \left( \delta^{AB} \delta^{CD} \frac{\partial^4}{\partial w_{s(\ell)}^A \partial \bar{w}_{s(\ell)}^B \partial w_{t(\ell)}^C \partial \bar{w}_{t(\ell)}^D }  \right)^{2j_\ell} \, .
\ee

\subsection{Special classes of squeezed vacua}

For special choices of the squeezing matrix $\gamma \in \mathcal{D}$, the squeezed vacuum $\ket{\Gamma,\gamma}$ correspond to a simple superposition of the coherent states discussed in Section \ref{sec:coherent}. This is true, in particular, for local squeezing matrices. In general, the coefficients $\gamma_{ij}^{AB}$ of the matrix $\gamma$ couple oscillators $a_i^A,a_j^B$ at arbitrary seeds $i,j$ of the graph $\Gamma$. When the coefficients $\gamma_{ij}^{AB}$ are nonzero only for pairs of seeds living in a common local patch of the graph $\Gamma$, we say that $\gamma$ is a local squeezing matrix. We shall consider two classes of local squeezing matrices. If $\gamma$ only couples seeds at the same node, we call it a nodewise squeezing matrix. If $\gamma$ only couples seeds at the same link, we call it a linkwise squeezing matrix. For these examples, the loop expansion \eqref{eq:loop-expansion-squezed} assumes particularly simple forms.

A linkwise squeezing matrix $\gamma_l(z)$ is defined as:
\be
[\gamma_l(z)]_{ij}^{AB} =
\begin{cases}
\lambda_\ell  z^A_i z^B_j  & \text{if } i,j \in \ell \text{ and } i\neq j\, , \\
  0 & \text{otherwise} \, ,
 \end{cases}
\ee
where $z^A_i \in \mathbb{C}^2$ are spinors attached to the seeds of $\Gamma$, and $\lambda_\ell$ is a real number associated with the link $\ell$. By construction, the squeezed state $\ket{\gamma_l(z)} \in \mathcal{H}_\mathcal{S}$ satisfies all the link constraints. The projection to the space of physical states is then implemented by the node projectors, yielding:
\begin{align}	
\ket{\Gamma,\gamma_l(z)} &= \left( \prod_n P_n \right) \exp\left(\frac{1}{2}\, [\gamma_l(z)]_{AB}^{ij}\, F^{AB}_{ij}{}^\dagger\right) \ket{0} \nonumber \\
	&= \sum_{j_\ell} \left(\prod_\ell \lambda_\ell^{2j_\ell} \right) \ket{\{j_i,z_i\}} \, .
\end{align}
We see that the state is a superposition of Livine-Speziale coherent states weighted by the product of powers of the parameter $\lambda_\ell$ at each link. The LS states are semiclassical states of the intrinsic geometry with well-defined areas fixed by the spins $j_\ell$. We have now an orthogonal superposition of such semiclassical states including arbitrary spins $j_\ell$. Instead of being peaked with minimal uncertainty at some classical configuration, the linkwise squeezed vacuum describe locally a mixture of semiclassical states with variable areas.

The loop expansion of $\ket{\Gamma,\gamma_l(z)}$ can be determined explicitly. The integration \eqref{eq:mu-Phi} giving the expansion coefficients $\mu_\Phi(\gamma)$ factorizes over the links of the graph for $\gamma_l(z)$, allowing us to perform the integration. We find the simple result:
\be
\ket{\Gamma,\gamma_l(z)} = \sum_\Phi Z_\Phi \frac{ \prod_\ell \lambda_\ell^{2j_\ell}}{\prod_n (J_n+1)!}   F_\Phi^\dagger \ket{0}\, ,
\label{eq:linkwise-loop-expansion}
\ee
where the holomorphic function $Z_\Phi$ is computed for the spinors $z_i^A$ of the squeezing matrix $\gamma_l(z)$.

A nodewise squeezing matrix $\gamma_0(z)$ is defined as:
\be
[\gamma_0(z)]_{ij}^{AB} =
\begin{cases}
\epsilon^{AB} \epsilon_{CD} z_i^C z^D_j  & \text{if } i,j \in n \, , \\
  0 & \text{otherwise} \, ,
 \end{cases}
 \label{eq:nodewise-gamma}
\ee
and is specified by the set of spinors $z_i^A$. The squeezing matrix $\gamma_0(z)$ only couples pairs of seeds $i\neq j$ at the same node. The requirement that $\gamma_0(z)$ lies in the Siegel unit disk $\mathcal{D}$ imposes the restriction  $0\leq \lambda_n<1$ at each node, where  $\lambda_n\equiv  \sum_{i\in n}|\vec{v}(z_i)|$. The properties of the nodewise squeezed vacuum $\ket{\gamma_0(z)} \in \mathcal{H}_\mathcal{S}$ in the full bosonic Hilbert space have been discussed in \cite{Bianchi:2016}. The state is a tensor product over nodes,
\begin{align}
\ket{\gamma_0(z)} &= \exp\left(\frac{1}{2}\, [\gamma_0(z)]_{AB}^{ij}\, F^{AB}_{ij}{}^\dagger\right) \ket{0} \\
	&= \,\bigotimes_{n\in\Gamma} \sum_{J_n=0,1,2,\ldots}\!\!\sqrt{J_n+1}\;{\lambda_n}^{J_n}\;|J_n,\{\hat{z}_i\}\rangle_{{\!}_n}\, ,
\label{eq:nodewise}
\end{align}
where we normalized the spinors $\hat{z}^i\equiv z^i/\sqrt{\sum_{i\in n}|\vec{v}(z_i)|}$ so that $\sum_{i\in n}|\vec{v}(\hat{z}_i)|=1$. This choice ensures that the $U(N)$ coherent intertwiners $\ket{J_n,\{\hat{z}_i\}}$ are normalized to $1$. Then the norm of the nodewise squeezed state is given by:
\be
\langle\gamma_0(z)|\gamma_0(z)\rangle=\prod_n (1-\lambda_n^2)^{-2} \, .
\ee
The probability of finding a total spin $J_n$ at a node reads $p(J_n)=(1-\lambda_n^2)^2\,(J_n+1)\lambda_n^{\,2J_n}$. The $U(N)$ coherent intertwiners are semiclassical states representing a region of space with a total boundary area $J_n$. The nodewise squeezed vacuum $\ket{\gamma_0(z)}$ is an orthogonal superposition of such states with a distribution of probabilities $p(J_n)$ close to a thermal distribution with temperature $2 \log 1/\lambda_n$.

The loop expansion of the nodewise squeezed state $\ket{\Gamma,\gamma_0(z)} = P_\Gamma \ket{\gamma_0(z)} \in \mathcal{H}_\Gamma$ can be determined using Eq.~\eqref{eq:loop-expansion-squezed}. The integral \eqref{eq:mu-Phi} defining the expansion coefficients $\mu_\Phi(\gamma)$ now factorizes over the nodes of the graph, and we find the simple formula:
\be
|\Gamma,\gamma_0(z)\rangle=\sum_\Phi \frac{Z_\Phi}{\prod_\ell (2j_\ell)!}  \,  F^\dagger_\Phi|0\rangle.
\label{eq:nodewise-loop-expansion}
\ee
The local structure of the squeezing matrix leads again to a compact expression for the loop expansion of the state.

Let us now consider a slightly more general class of squeezed vacua in which small nonlocal components are allowed in the squeezing matrix and treated as perturbations around a local squeezing matrix. We introduce for that a squeezing matrix $\gamma_1$ of the form:
\be
[\gamma_1]_{ij}^{AB} = \epsilon^{AB} \gamma_{ij} \, , \qquad \gamma_{ij} \in \mathbb{C} \, , 
\label{eq:perturbed-nodewise}
\ee
and decompose $\gamma_{ij}$ into a sum of diagonal and purely off-diagonal components:
\be
\gamma_{ij} = \gamma_{ij}^{(D)} + \varepsilon \delta \gamma_{ij} \, .
\label{eq:fact-pert-nodewise}
\ee
The loop expansion of $\ket{\Gamma,\gamma_1}$ can be determined to first order in $\delta \gamma_{ij}$. In order to describe it, let us first define:
\be
\gamma_\Phi \equiv
\prod_{\alpha\in\Phi}\Big(\prod_{\langle i,j\rangle\in \alpha} \gamma_{ij} \Big)^{m_\alpha}\,.
\label{eq:gamma-phi}
\ee
Moreover, for a product of two $\gamma_{ij}$ matrix elements, we define a braiding
\begin{equation}
B(\gamma_{ij} \gamma_{k\ell}) = 
\begin{cases}
\gamma_{ik} \gamma_{j\ell} - \gamma_{i \ell}\gamma_{jk} & \text{if } (i,j) \in n \text{ and } (k,\ell) \in n' \text{ for nodes } n \ne n' \, ,\\
0 & \text{otherwise} \, .
\end{cases}
\label{eq:braiding}
\end{equation}
Note that this operation probes the off-diagonal elements of $\gamma_{ij}$. We extend this operation to higher order monomials in $\gamma_{ij}$ via
\begin{equation}
B(\gamma ... \gamma ) = \sum_{\text{pairs } (\gamma_{ij}, \gamma_{k\ell})} \gamma ... B(\gamma_{ij} \gamma_{k\ell}) ... \gamma
\label{eq:braid-extend}
\end{equation}
The loop expansion of the squeezed vacuum $\ket{\Gamma,\gamma_1}$ is then given by
\begin{align}
\ket{\Gamma,\gamma_1} = \sum_\Phi \frac{1}{ \prod_\ell (2 j_\ell)!}\left[\gamma_\Phi+ \frac{1}{2} B(\gamma_\Phi) + O(\epsilon^3)\right] F^\dagger_\Phi \ket{0} \, ,
\label{eq:cycle-decomp}
\end{align}
as proved in Appendix \ref{sec:perturbed-nodewise}. Note that if we set $\varepsilon=0$ and $\gamma_{ij}^{(D)}=\epsilon_{CD} z_i^C z_J^D$, the squeezing matrix $\gamma_1$ reduces to the nodewise squeezing matrix \eqref{eq:nodewise-gamma}. In this case, the braiding term vanishes and $\gamma_\Phi=Z_\Phi$, so that we recover Eq.~\eqref{eq:nodewise-loop-expansion}. A squeezing matrix of the form $\gamma_1$ has been studied in \cite{Bianchi:2016} for a graph $\Gamma$ of cubic structure, and shown to define a squeezed state with long range spin-spin correlations that decay as the inverse of the squared distance, reproducing the typical behavior of correlations for fluctuations of massless quantum fields in a classical background. Such a state is there proposed as a candidate for the description of the vacuum of the graviton in a background space determined by the diagonal part of the squeezing matrix.

\subsection{Heat kernel states}

The physical states $\ket{\psi} \in \mathcal{H}_\Gamma$ of loop quantum gravity on a graph $\Gamma$ describe the quantum geometry of a three-dimensional slice of spacetime, including the intrinsic and extrinsic geometry. The intrinsic geometry is encoded in the areas of faces and dihedral angles in the dual lattice $\Gamma^*$, determined by the spins $j_\ell$, while the extrinsic curvature is encoded in the holonomies $h_\ell$. The coherent states discussed in Section \ref{sec:coherent} are peaked in classical configurations of the intrinsic geometry determined by the spinors $z_i$ involved in their construction. Semiclassical states known as heat kernel states that are peaked on both the intrinsic and extrinsic geometries have been constructed in \cite{Thiemann:2000bw,Thiemann:2002vj}, and extensively discussed in the literature \cite{Thiemann:2000ca,Thiemann:2000bx,Thiemann:2000by,Bahr:2007xa,Bahr:2007xn}. The classical phase space of loop quantum gravity on a graph $\Gamma$ is the space of twisted geometries on the dual graph \cite{Freidel:2010aq}. A heat kernel state is naturally associated with a classical twisted geometry on $\Gamma^*$, and thus labeled by a point in the phase space of the theory \cite{Bianchi:2009ky}.

A heat kernel state $\ket{\Gamma,H_\ell,t_\ell} \in \mathcal{H}_\Gamma$ is characterized by a choice of an element $H_\ell$ of the group $SL(2,\mathbb{C})$ and a real parameter $t_\ell$ for each link $\ell$ of the graph $\Gamma$. One first defines a heat kernel state $\ket{H_\ell,t_\ell} \in \mathcal{H}_\mathcal{S}$ in the full bosonic space as a tensor product over links:
\be
\ket{H_\ell,t_\ell} = \bigotimes_\ell \sum_{j_\ell} (2j_\ell+1) e^{-t_\ell j_\ell(j_\ell+1)} \left[D^{(j_\ell)}(H_\ell)\right]_{mn} \ket{j_\ell,m,n}\, ,
\ee
which satisfies all link constraints. The state $\ket{\Gamma,H_\ell,t_\ell}$ is then obtained by projecting to the space of physical states,
\be
\ket{\Gamma,H_\ell,t_\ell} = P_\Gamma \ket{H_\ell,t_\ell} \, .
\ee
The scalar product of a heat kernel state with a multiloop state $F_\Phi^\dagger|0\rangle$ defines a function $\mu_\Phi(H_\ell,t_\ell)$ that can be expressed as a complex integral,
\begin{align}
\mu_\Phi(H_\ell,t_\ell) &\equiv \langle 0|F_\Phi \ket{\Gamma,H_\ell,t_\ell}\\
&= \left[ \prod_\ell (2j_\ell+1) e^{-t_\ell j_\ell(j_\ell+1)} \right]
\int \frac{d^{4L}z \, d^{4L}\bar{z}}{\pi^{4L}} \, Z_\Phi\;
e^{-z_i^A \bar{z}^i_A+\frac{1}{2} [\gamma(H)]_{ij}^{AB} \bar{z}_A^i \bar{z}_B^j}  
\label{eq:mu-H-t}
\end{align}
where we introduced the symmetric matrix $\gamma(H) \in \textrm{Mat}(4L,\mathbb{C})$ with components:
\be
[\gamma(H)]_{ij}^{AB} = 
\begin{cases}
[H_\ell]^{AB}   & \text{if } i=s(\ell) \text{ and } j=t(\ell) \, , \\
  0 & \text{if } (i,j) \text{ is not a link} \, ,
 \end{cases}
 \label{eq:heat-kernel-gamma}
\ee
The integral in Eq.~\eqref{eq:mu-H-t} evaluate to $\prod_\ell (2j_\ell)! (\mathbf{\epsilon} \cdot \mathbf{H})$, where $(\mathbf{\epsilon} \cdot \mathbf{H})$ represents the contraction of the $\epsilon_{AB}$ tensors associated with the wedges of $\Phi$ with the $SL(2,\mathbb{C})$ elements $H_\ell^{AB}$ at the links traversed by $\Phi$. Using the representation \eqref{eq:loop-expansion} of the projector $P_\Gamma$ in \eqref{eq:squeezed-lqg} and inserting a coherent state resolution of the identity in the resulting expression, we find the representation of $\ket{\Gamma,H_\ell,t_\ell}$ as a superposition of multiloop excitations:
\begin{equation}
\ket{\Gamma,H_\ell,t_\ell} = \sum_\Phi \frac{ \left[\prod_\ell (2j_\ell+1) e^{-t_\ell j_\ell(j_\ell+1)} \right] (\mathbf{\epsilon} \cdot \mathbf{H}) }{\prod_n(J_n+1)!}\;F_\Phi^\dagger|0\rangle.
\label{eq:loop-expansion-heat-kernel}
\end{equation}
Note that we can use the matrix $\gamma(H)$ as a squeezing matrix and represent the heat kernel states in the form:
\be
\ket{H_\ell,t_\ell} = \left. \left[\prod_\ell \left( 2I_{\ell}+1\right ) e^{-t_\ell I_\ell (I_\ell+1) } \right] \exp\left(\frac{1}{2}\,\gamma^{ij}_{AB} \, F^{AB}_{ij}{}^\dagger \right) \ket{0} \right|_{\gamma \to \gamma(H)}\, .
\label{eq:heat-kernel-squeezed}
\ee
Note that $\gamma(H)$ does not belong to the Siegel unit disk $\mathcal{D}$ in general, but by acting with the operator in square brackets on the power series of the exponential, we obtain an expression that is finite when computed at $\gamma(H)$.


\section{Generating function for squeezed vacua}
\label{sec:generating-function}

A generic state $\ket{\psi} \in \mathcal{H}_\Gamma$ naturally decomposes into a sum of orthogonal components with fixed spins $j_\ell$. In the case of projected squeezed vacua, we can write:
\be
\ket{\Gamma,\gamma} = \sum_{j_\ell} \ket{\Gamma, \gamma,j_\ell} \, , \qquad \ket{\Gamma, \gamma,j_\ell} =\mathcal{P}_{\{j_\ell\}} \ket{\Gamma,\gamma} \, ,
\label{eq:generating-j-expansion}
\ee
where $\mathcal{P}_{\{j_\ell\}}$ is the projector onto the space of states with spin configuration $\{j_\ell\}$. We wish to show that the orthogonal pieces $\ket{\Gamma, \gamma,j_\ell}$ can be written in terms of partial derivatives of a single generating function
\be
G(\gamma,x) = \det \left(1-\gamma^{ij}_{AB} \epsilon^{BC} F^\dagger_{jk}(x) \right)^{-1/2}  \, ,
\label{eq:generating-function}
\ee
where
\be
F^\dagger_{ij}(x) \equiv \begin{cases}
F_{ij}^\dagger x_i x_j 	& \text{for } i,j \in n \, , \\
0					& \text{otherwise} \, .
\end{cases}
\ee
Here we have assigned real variables $x_i \in \mathbb{R}$ to the seeds $i$ of the graph. The orthogonal projection of a squeezed vacuum onto the subspace with spins $j_\ell$ is obtained by taking the $2 j_\ell$-th derivative of the generating function $G(\gamma,x)$ with respect to the seeds at the link $\ell$:
\be
\ket{\Gamma, \gamma,j_\ell} = \frac{1}{\prod_n (J_n+1)!} \prod_\ell \frac{1}{(2 j_\ell)!^2}  \left( \frac{\partial^2}{\partial x_{s(\ell)} \partial x_{t(\ell)}} \right)^{2j_\ell} G(\gamma,x)\bigg|_{x=0} \ket{0} \, .
\label{eq:F-x}
\ee

This result is obtained immediately by using the techniques developed for the derivation of the loop expansion in Section \ref{sec:loop-expansion}. From Eq.~\eqref{formal_proj1}, we have:
\begin{align}
& \sum_{J_n} \left[\prod_n(J_n+1)! \, \mathcal{P}_{J_n} \right] \ket{\gamma} \notag \\
& \qquad \qquad =  \normord{ \exp\left(\sum_n \sum_{i,j \in n} \frac{1}{2} F_{ij}^\dagger F_{ij} - 2 \sum_i I_i\right)} \ket{\gamma} \notag \\
& \qquad \qquad = \int \frac{d^{4L}z \, d^{4L}\bar{z}}{\pi^{4L}} \exp\left( -z^A_i \bar{z}^i_A + \sum_n \sum_{i,j\in n} \frac{1}{2}z^i_A(\epsilon^{AB}F^\dagger_{ij})z_B^j  + \frac{1}{2}\bar{z}_A^i \gamma_{ij}^{AB} \bar{z}_B^j \right) \ket{0} \label{eq:G-integral-rep}\\
& \qquad \qquad = G(\gamma,x) |_{x=1} \ket{0}  \, .
\end{align}
We first introduced a coherent state resolution of the identity and then computed the resulting gaussian integral. Note that the integral representation \eqref{eq:G-integral-rep} of the generating function $G(\gamma,x)$ is valid for any $x$ provided that one replaces $F_{ij}^\dagger \to F_{ij}^\dagger(x)$. The dummy variables $x_i$ introduced in Eq.~\eqref{eq:F-x} are used to keep track of the number of excitations at a given seed $i$. The projection to the space of fixed spins is then implemented by:
\be
\mathcal{P}_{\{j_\ell\}} \sum_{J_n} \left[\prod_n(J_n+1)! \, \mathcal{P}_{J_n} \right] \ket{\gamma}  = \prod_\ell \frac{1}{(2 j_\ell)!^2}  \left( \frac{\partial^2}{\partial x_{s(\ell)} \partial x_{t(\ell)}} \right)^{2j_\ell} G(\gamma,x)\bigg|_{x=0} \ket{0} \, .
\ee
The expression on the left-hand side of this equation corresponds to $\mathcal{P}_{\{j_\ell\}} \ket{\Gamma,\gamma}$ except for the combinatorial factors. Dividing both sides of the expression by $\prod_n (J_n+1)!$ we arrive at \eqref{eq:F-x}.

In the loop representation discussed in Section \ref{sec:squeezed}, the explicit representation of a squeezed vacuum $\ket{\Gamma,\gamma}$ requires the computation of the amplitudes $\mu_\Phi(\gamma)$ of the loop excitations $\ket{\Phi}$, for all multiloops $\Phi$. Each such coefficient is given by a complex integral \eqref{eq:mu-Phi}. In the alternative representation provided by the generating function $G(\gamma,x)$, the explicit description of a state is reduced to the computation of a single determinant, Eq.~\eqref{eq:generating-function}. In general, this determinant is too complex to be computed in closed form, but in situations where it can be determined, the generating function formalism offers an efficient method for the study of the quantum geometry of squeezed vacua. A simple example is discussed in the next section.

\subsection{Generating function for two loop states}

Let the graph $\Gamma$ be the union of two isolated loops $\alpha_1,\alpha_2$ formed by the links $\ell_1=\{1,2\},\ell_2=\{3,4\}$. Since the graph has four seeds, the corresponding Schwinger model has eight oscillators $a_i^A$, $i=1,\dots,4$. Consider the squeezing matrix:
\begin{equation}
\gamma^{ij}_{AB} = 
\left(
\begin{array}{cc}
\;\beta_1 \;\epsilon_{AB}\, \epsilon^{i_1j_1}\;  &\; \lambda\; \delta_{AB} \,\delta^{i_1 j_2}     \\[1em]
\;\lambda \;\delta_{AB}\, \delta^{i_2 j_1}\;  &\;  \beta_2  \;\epsilon_{AB}\, \epsilon^{i_2j_2}
\end{array}
\right) \, ,
\label{eq:gamma-two-loop}
\end{equation}
where $i_1,j_1=1,2$ and $i_2,j_2=3,4$. The matrix $\gamma$ is presented in block form with respect to the loops. The generating function $G(\gamma,x)$ can be exactly determined:
\be
G(\gamma,x) = \sum_{j_1,j_2} (2j_1+1)(2j_2+1)
(\beta_1 x_1 x_2 F_{12}^\dagger)^{\,2j_1} (\beta_2 x_3 x_4 F_{34}^\dagger)^{\,2j_2}\;{}_2 F_1\Big(\!\!-2j_1,-2j_2,\,2,\frac{\lambda^2}{\beta_1 \beta_2}\Big) \, . 
\ee
An explicit expansion of the squeezed vacuum $\ket{\Gamma,\gamma}$ in components with well-defined spins is now obtained from Eqs.~\eqref{eq:generating-j-expansion} and \eqref{eq:F-x}:
\be
\ket{\Gamma,\gamma} =\mathcal{N}\sum_{j_1,j_2}\sqrt{(2j_1+1)(2j_2+1)}
\beta_1^{\,2j_1}\beta_2^{\,2j_2}\;{}_2 F_1\Big(\!\!-2j_1,-2j_2,\,2,\frac{\lambda^2}{\beta_1 \beta_2}\Big)\,|\alpha_1,j_1\rangle|\alpha_2,j_2\rangle
\label{eq:two-loop-state}
\ee
where ${}_2 F_1(a,b,c,z)$ is the Gauss hypergeometric function and normalized loop states with fixed spins were introduced:
\be
\ket{\alpha_1,j_1} = \frac{1}{\sqrt{(2 j_1)! (2j_1+1)!}} \bigl(F_{12}^\dagger\bigr)^{2j_1} \ket{0} \, ,
\ee
and similarly for the loop $\alpha_2$. 

Let us consider some special cases. For $\lambda=0$, the hypergeometric function evaluates to $1$, and the state becomes separable. In this case, the squeezing matrix \eqref{eq:gamma-two-loop} is block-diagonal, and excitations are created independently in the two loops. Correlations can be introduced by switching on the off-diagonal elements of $\gamma$.

For a purely off-diagonal squeezing matrix, $\beta_1,\beta_2 \to 0$, we have:
\be
\lim_{\beta_1,\beta_2 \to 0} \beta_1^{\,2j_1}\beta_2^{\,2j_2}\;{}_2 F_1\Big(\!\!-2j_1,-2j_2,\,2,\frac{\lambda^2}{\beta_1 \beta_2}\Big) = \delta_{j_1,j_2} \frac{\lambda^{4j_1}}{2j_1+1} \, ,
\ee
leading after normalization to
\be
\ket{\Gamma,\gamma} = \sqrt{1-|\lambda|^4} \sum_j \lambda^{4j} \ket{\alpha_1,j}\ket{\alpha_2,j} \, .
\ee
We see that the states of the two loops are perfectly correlated. Moreover, the reduced density matrix $\rho_1$ describing the subsystem associated with the loop $\alpha_1$ is given by a thermal distribution 
\be
\rho_1 \propto e^{- \mu H_{E}} \, ,
\ee
with inverse temperature $\mu= -8 \log |\lambda|$ and entanglement Hamiltonian $H_E=I_1$.

\begin{figure}[t]
\centering
\includegraphics[height=16em]{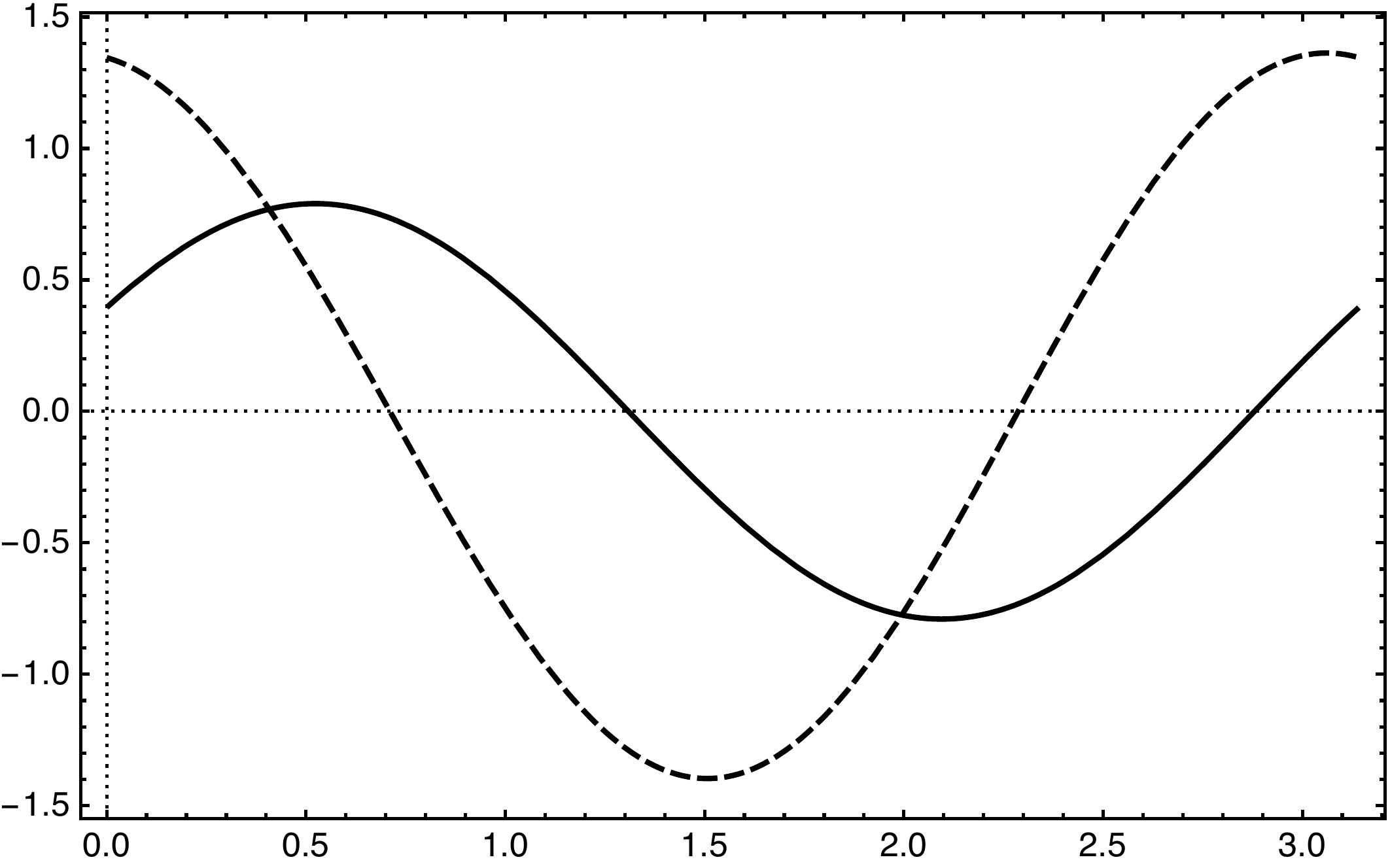}
\caption{Correlation functions $\mathcal{C}_{I}(\phi)=\langle I_1 \,I_2\rangle-\langle I_1\rangle \langle I_2\rangle$ (\emph{solid line}) and $\mathcal{C}_{W}(\phi)=\langle W_1 W_2\rangle-\langle W_1\rangle \langle W_2\rangle$ (\emph{dashed line}) in a squeezed vacuum, Eq.~\eqref{eq:two-loop-state}. The squeezing parameters $\beta_1=\beta_2=r e^{i \theta}, \lambda=\varepsilon e^{i \phi}$ are set to $r=0.5$, $\theta=\pi/6$, $\varepsilon=0.1$, with $\phi$ varying in the range $[0,\pi]$. The correlations functions vanish as $\sim \varepsilon^2$ for $\varepsilon\to 0 $. The ordinate axis is represented in units $\varepsilon^{-2}$.}
\label{fig:plot-correlations}
\end{figure}

In the presence of both diagonal and off-diagonal components, with an off-diagonal part $\lambda = \varepsilon e^{i \phi}$, where $\varepsilon$ is taken to be small, we have after normalization:
\be
\ket{\Gamma,\gamma} = \frac{1}{\kappa} \sum_{j_1,j_2} \sqrt{(2 j_1+1)(2 j_2+1)}\beta_1^{2 j_1} \beta_2^{2 j_2} \left[ 1+ \frac{\varepsilon^2}{2}  \left( 4 j_1 j_2 \frac{e^{2 i \phi}}{\alpha \beta} - \kappa \sigma \right)\right] \ket{\alpha_1,j}\ket{\alpha_2,j} + O(\epsilon^3) \, ,
\label{eq:case-3}
\ee
where we have defined
\begin{align}
\kappa &= \frac{1}{(1-|\beta_1|^2)(1-|\beta_2|^2)} \, ,\\
\sigma &= 4 \,\text{Re}(\beta_1 \beta_2 e^{-2 i \phi}).
\end{align}
We can use Eq.~\eqref{eq:case-3} to determine the correlation functions:
\begin{align}
\mathcal{C}_{I}&\equiv \langle I_1 \, I_2\rangle-\langle I_1\rangle\langle I_2\rangle\\[.5em]
\mathcal{C}_{W}&\equiv \langle W_1 W_2\rangle-\langle W_1\rangle\langle W_2\rangle \, 
\end{align}
with the Wilson loop operators $W$ defined in Eq.~\eqref{eq:wilson-loop}. In Fig.~(\ref{fig:plot-correlations}) we plot the correlations as a function of $\phi$ for $\beta_1=\beta_2=r e^{i \theta}$ to illustrate the behavior of these quantities. By allowing for variations of all squeezing parameters, the correlations $\mathcal{C}_{I}$ and $\mathcal{C}_{W}$ can be varied independently.


\section{Conclusion}
\label{sec:conclusion}

We introduced a new basis of loop states for the Hilbert space $\mathcal{H}_\Gamma$ of loop quantum gravity on a graph $\Gamma$ by making full use of bosonic techniques developed in the context of the spinor formalism \cite{Girelli:2005ii,Freidel:2010tt,Dupuis:2010iq,Borja:2010rc,Bonzom:2012bn,Livine:2011gp,Livine:2011zz,Livine:2013wmq}. We showed that the overcompleteness of the loop basis that has historically prevented its application in practical problems can be dealt with by working with normal-ordered versions of the Wilson loop operators, naturally available in the bosonic representation. This simple modification leads to a vast reduction in the number of loop states to be considered, owing to the fact that normal-ordered Wilson loop operators vanish for loops with tails of the form $\gamma\circ\gamma^{-1}$. This blocks the possibility of trivially deforming the loops to produce new states, preventing their excessive proliferation and in this way keeping the new basis at a manageable degree of overcompleteness. The new loop states satisfy a reduced set of local Pl\"ucker identities at the nodes of $\Gamma$ that replace the usual Mandelstam and retracing identities \cite{Gambini:1996ik,Rovelli:1995ac}. A resolution of the identity was constructed in the new loop basis and used to determine explicit loop expansions of a large class of states.

We started our construction with the definition of the new loop basis. In the bosonic representation, $\mathcal{H}_\Gamma$ corresponds to the space of solutions to the area matching and Gauss constraints in a bosonic Hilbert space $\mathcal{H}_{\mathcal{S}}$ of harmonic oscillators living on the graph $\Gamma$. The space $\mathcal{H}_\Gamma$ is equipped with a representation of the holonomy-flux algebra \cite{Livine:2013wmq}. We first fixed an ordering ambiguity in the holonomy operators by requiring their eigenstates to correspond to delta functions in the standard holonomy representation, leading to the symmetric ordering given in Eq.~\eqref{eq:hl}. Wilson multiloop operators $W_\Phi$ were then defined in the usual way, and we introduced operators $F^\dagger_\Phi$ such that $\normord{W_\Phi}\ket{0}=F_\Phi^\dagger \ket{0}$. For any multiloop $\Phi$, the operator $F_\Phi^\dagger$ is written in Eq.~\eqref{eq:multiloop-annihilation} as the product of invariant creation operators of the $U(N)$ formalism \cite{Freidel:2010tt,Borja:2010rc} living at the nodes of $\Phi$. The new loop states are defined as $\ket{\Phi}=F_\Phi^\dagger \ket{0} \in \mathcal{H}_\Gamma$.

The projection operator $P_\Gamma:\mathcal{H}_{\mathcal{S}} \to \mathcal{H}_\Gamma$ is a product of projectors $P_\ell$ and $P_n$ associated with links $\ell$ and nodes $n$ of $\Gamma$, which solve the area matching and Gauss constraints, respectively. We derived explicit formulas for the local projectors in two alternative forms: as normal-ordered Bessel functions of bosonic operators, Eqs.~\eqref{eq:cl-bosonic} and \eqref{eq:cn-bosonic}, and in a diagonal coherent state representation, Eqs.~\eqref{eq:cl-coherent-1} and \eqref{eq:cn-coherent-1}. The normal-ordered representation of the local projectors is the basis for the derivation of the loop expansion \eqref{eq:Ploop} of $P_\Gamma$. A compact proof of the expansion was given in Section \ref{subsec:multiloop}. Since it requires the formal manipulation of divergent operators, we also presented a second, more laborious derivation involving only well-defined operators in order to support the result. The resulting loop expansion involves only non-repeating multiloops, formed by loops $\alpha_i$ which do not admit a representation of the form $\alpha=\beta^n$ with $n\neq1$, and automatically discards loops with trivial tails of the form $\gamma\circ\gamma^{-1}$, for which $\ket{\Phi}$ vanishes. Redundancies in the expansion can be further eliminated by introducing classes of equivalence of multiloops symmetrized along links, yielding a picture closely related to the spin network basis, but with intertwiner spaces described by overcomplete bases labeled by segments of curves crossing the nodes. This gives an alternative representation of $P_\Gamma$ as a resolution of the identity in a basis of symmetrized loop states.

The projector $P_\Gamma$ was then applied for a variety of states. We focused on the familiar classes of semiclassical states of loop quantum gravity, including coherent, squeezed and heat kernel states. These states are naturally defined in the bosonic space $\mathcal{H}_\mathcal{S}$ and then projected down to $\mathcal{H}_\Gamma$. We showed that the projection can be efficiently implemented using our loop expansion of $P_\Gamma$, yielding concrete representations of such states as superpositions of multiloop excitations $\ket{\Phi}$ in $\mathcal{H}_\Gamma$. The loop expansion of coherent states is given in Eq.~\eqref{eq:loop-expansion-coherent}, and that of heat kernel states in Eq.~\eqref{eq:loop-expansion-heat-kernel}. For the case of squeezed states, the loop expansion can be computed in closed form for squeezing matrices that are local with respect to links, Eq.~\eqref{eq:linkwise-loop-expansion}, or nodes, Eq.~\eqref{eq:nodewise-loop-expansion}, and in the presence of small off-diagonal perturbations, Eq.~\eqref{eq:cycle-decomp}.

The loop expansion of coherent, heat kernel states and locally squeezed states is such that the amplitude $c_\Phi$ of a multiloop excitation $\ket{\Phi}$ is a product of local weights picked up at the links and nodes traversed by the loop. The factorizable form of the loop amplitudes reflects the local nature of such states, which are defined as separable states in $\mathcal{H}_\mathcal{S}$ before being projected to $\mathcal{H}_\Gamma$. Such a factorization does not occur for a squeezing matrix with nonzero off-diagonal terms. As discussed in \cite{Bianchi:2016}, non-local terms in the squeezing matrix encode long-range correlations in the fluctuations of the geometry, suggesting a relation between the presence of long-range correlations and a non-factorizable form of the loop amplitudes $c_\Phi$. Such a relation can be established rigorously for states close to the Ashtekar-Lewandowski vacuum \cite{Bianchi:2016lrc}, and it is important to explore this correspondence further for other classes of states. 
 
We also constructed a generating function $G(\gamma,x)$ for squeezed vacua whose derivatives give the projections of the states onto the subspaces of fixed spins $j_\ell$. We applied this technique to the simple case of a graph $\Gamma$ formed by two disconnected loops. The generating function can be computed in closed form for this example and was applied to the calculation of average values and correlation functions for the spins and Wilson loops. Varying the parameters of the squeezing matrix, the correlation functions can be tuned at will, and we described how separable and locally thermal states can be obtained. This example illustrates in a simple context how our bosonic techniques can be applied to the manipulation of quantum correlations in fluctuations of the geometry.

The mathematical tools introduced in this paper were developed with the objective of identifying the semiclassical regime of loop quantum gravity by the characterization of correlations in the fluctuations of the quantum geometry. The idea that classical geometry emerges from the structure of correlations of a quantum state has been studied along several lines recently \cite{Bianchi:2012ev,VanRaamsdonk:2010pw,Czech:2012bh,Kempf:2009us,Saravani:2015moa,Cao:2016mst,Chirco:2014naa}. In the reconstruction of spacetime proposed in \cite{Bianchi:2012ev}, an area law for the entanglement entropy plays a central role (see also \cite{Cao:2016mst}). In \cite{Chirco:2014naa}, the density matrix for a finite region of space is required to be a KMS state in order to reproduce the general form of the vacuum of quantum field theories. The construction of states with prescribed correlations is a key issue for the concrete implementation of these ideas in loop quantum gravity. In \cite{Bianchi:2016lrc} we proposed the application of squeezed vacua for this purpose. The mean geometry can be encoded in the local, diagonal coefficients of the squeezing matrix $\gamma$, while correlations are introduced as off-diagonal perturbations of $\gamma$. Here we computed the loop expansion of such perturbed squeezed states, providing a concrete representation useful for a more detailed study of their correlation functions. The explicit form of the known semiclassical states for the mean geometry in the loop basis makes the loop representation a convenient framework for this approach to the analysis of the classical limit of loop quantum gravity.


\begin{acknowledgments}
We thank Abhay Ashtekar, Wolfgang Wieland and Bekir Bayta\c{s} for numerous discussions on coherent and squeezed states. The work of EB is supported by the NSF grants PHY-1404204. NY acknowledges support from CNPq, Brazil and the NSF grant PHY-1505411.
\end{acknowledgments}

\appendix

\section{From routings to multiloops}
\label{sec:count-routings}

In the routing representation, each link with spin $j_\ell$ corresponds to a collection of $2 j_\ell$ distinguishable strands. A routing is obtained by joining the endpoints of such strands at each node in such a way that only strands of distinct links are connected. Let us label the  strands as $\lambda_{\ell \mu}$, $\mu=1,\dots ,2j_\ell$, and represent their source and target endpoints as $s(\lambda_{\ell \mu})$ and $t(\lambda_{\ell \mu})$, respectively. (The orientation of the links is arbitrary and does not affect the results.) The strand endpoints can be labeled by an index $r$. Then a routing can be represented by a collection of wedges $w=\{i,j\}$\footnote{These are wedges connecting strand endpoints, not seeds.}. We define two kinds of transformations acting on routings. The first is defined for generic routings:
\begin{itemize}
\item \emph{Link permutations $P$.} Strands are permuted within links. A permutation $\pi_\ell \in S_{2j_\ell}$ acts on wedges connected to $\ell$ as:
\begin{align}
\{i,s(\lambda_{\ell ,\mu})\} &\mapsto \{i,s(\lambda_{\ell ,\pi(\mu)})\} \, , \nonumber \\
\{j,t(\lambda_{\ell ,\mu})\} &\mapsto \{j,t(\lambda_{\ell ,\pi(\mu)})\} \, .
\end{align}
This operation does not change the multiloop associate with the routing. A generic link permutation $P$ is a composition of permutations $\pi_\ell$ over an arbitrary family of links $\ell$. Any two routings $R,R'$ of the same multiloop are related by some link permutation $P$.
\end{itemize}
The second operation is defined only for routings of non-repeating multiloops:
\begin{itemize}
\item \emph{Loop braidings $B$.} Let $\ell$ be the first link of the loop $\alpha_1$ in the non-repeating multiloop $\Phi=\{ \alpha_1^{N_1}, \alpha_2^{N_2}, \dots \}$ associated with some routing $R$.\footnote{Note that the elementary loops $\alpha_i$ have a starting point and are oriented.} For each of the $N_1$ copies of $\alpha_1$, there is a strand $\lambda_{\ell ,a}$ of $\ell$ traversed by the first link of $\alpha_i$. Permuting the sources of such strands among themselves with some $\pi_b \in S_{N_1}$, the wedges attached to them transform as:
\be
\{i,s(\lambda_{\ell ,a})\} \mapsto \{i,s(\lambda_{\ell ,\pi(a)})\} \, .
\ee
Similar transformations are defined for all $\alpha_i$. A generic braiding $B$ is a composition of loop braidings $\pi_b$ for all kinds of elementary loops $\alpha_i$ in $\Phi$. Any nontrivial loop braiding changes the multiloop associated with the routing, since it creates a repeating multiloop. All multiloops $\tilde{\Phi}$ in the equivalence class of $\Phi$ can be generated in this way. 
\end{itemize} 

The loop braidings have the following property: if two non-repeating routings $R,R'$ of $\Phi$ are mapped by braidings $\pi_b,\pi_b'$ into a common image $\tilde{R}$, then they must be the same:
\be
B(R) = B'(R') \implies R=R' \, .
\label{eq:uniq}
\ee
In order to see this, first note that braidings $\pi_b$ of a loop $\alpha_i$ act only on wedges at a single node $n$, the source of the loop $\alpha_i$. The operation is local. Therefore, if $\pi_b(R) = \pi_b'(R')$, then $R$ and $R'$ must be identical at all nodes except for $n$, where wedges traversed by the copies of $\alpha_i$ could differ. But since the routings are non-repeating, there is only one way to connect the strands of the copies of the $\alpha_i$ reaching $n$ with wedges at $n$. Hence, it must be $R=R'$. This argument can be applied independently for all elementary braidings $\pi_b^{(i)}$  of distinct elementary loops $\alpha_i$.

A second property of braidings is that any routing $R$ is the image of a non-repeating routing $R'$ under some braiding, $R= B(R')$. $R'$ is the non-repeating routing which is identical to $R$ at all nodes except at the sources of the first strands of each copy of the $\alpha_i$.

Now let us count how many routings are associated with a non-repeating multiloop $\Phi$. First choose some reference routing $R_0$ of $\Phi$. The group $\mathcal{P}$ of link permutations of $R$ has $\prod (2j_\ell)!$ elements. Not all of them produce distinct results, however, since whole loops can be permuted among themselves. Hence, the total number of distinct routings in the orbit $\mathcal{P}(R_0)$ is equal to $\prod (2j_\ell)!/\prod N_i!$. All such routings are non-repeating. Routings of repeating multiloops associated with $\Phi$ are now produced by the application of braidings. The number of braidings which can be applied to each $P (R_0)$ is equal to $\prod N_i!$. These always produce distinct results, as previously shown. Therefore, there are $\prod (2j_\ell)!$ distinct routings of the form $B\circ P(R_0)$. Now let $R$ be an arbitrary routing of a multiloop equivalent to $\Phi$. It can always be written as $B(R')$, where $R'$ is non-repeating. But $R'=P(R_0)$, for some $P$. Therefore, it must be of the form $R=B\circ P(R_0)$. We conclude that there are exactly $\prod (2j_\ell)!$ in the equivalence class of $\Phi$.

\section{Loop expansion of perturbed nodewise squeezed vacua}
\label{sec:perturbed-nodewise}

We wish to prove Eq.~\eqref{eq:cycle-decomp} describing the loop expansion of a perturbed nodewise squeezed vacuum associated with the squeezing matrix $\gamma_1$ defined in Eqs.~\eqref{eq:perturbed-nodewise} and \eqref{eq:fact-pert-nodewise}. We begin by stating two facts. Let $D_{ij}$ be the derivative operator
\begin{equation}
D_{ij} \equiv \epsilon^{AB} \frac{\partial^2}{\partial w^A_i \partial w^B_j}.
\end{equation}
We will call $D_{ij}$ mixed if $i$ and $j$ belong to different nodes, and unmixed otherwise. 

\begin{prop}\label{deriv_formula}
Let $ (i_1, j_1),...,(i_k,j_k)$ refer to pairs of seeds living at a common node. Then:
\begin{equation}
(D_{i_1j_1}...D_{i_kj_k})\exp\left(\frac{1}{2} \epsilon^{AB} F_{ij}^\dagger w^i_A w^j_B\right)\bigg|_{w=0} = (k+1)!\, F^\dagger_{i_1j_1}...F^\dagger_{i_kj_k}.
\end{equation}
\end{prop}
\noindent
One can prove this result via induction combined with the identity $F^\dagger_{ij}F^\dagger_{kl} =F^\dagger_{ik}F^\dagger_{jl}-F^\dagger_{il}F^\dagger_{jk}$. 
\begin{prop}\label{deriv_disentangle}
Let the indices $(i,j)$ refer to seeds at node $n$ and $(k,\ell)$ to seeds at $n'$ with $n\ne n'$. Then:
\begin{multline}
(D_{i_1k_1} D_{j_1\ell_1})(D_{i_2j_2}D_{i_3j_3}...)(D_{k_2\ell_2}D_{k_3\ell_3}...)\exp\left(\frac{1}{2} \epsilon^{AB} F_{ij}^\dagger w^i_A w^j_B \right)\bigg|_{w=0} \\
= \hspace{5pt} \frac{1}{2}(D_{i_1j_1}D_{i_2j_2}...)(D_{k_1\ell_1}D_{k_2\ell_2}...)\exp\left(\frac{1}{2} \epsilon^{AB} F_{ij}^\dagger w^i_A w^j_B \right)\bigg|_{w=0}.
\label{eq:jl-coefficient}
\end{multline}
\end{prop}
\noindent
Proposition (\ref{deriv_disentangle}) allows us to disentangle mixed derivatives that act across nodes. The resulting derivatives can then be evaluated using Proposition (\ref{deriv_formula}). (Since $F_{ij}^\dagger$ is block diagonal with respect to the local node Hilbert spaces, Proposition (\ref{deriv_formula}) can be applied to derivatives acting on multiple nodes, as long as the derivatives are not mixed.)

We now evaluate Eq.~\eqref{eq:squeezed-der-repr} for the squeezing matrix $\gamma_1$ of the form $[\gamma_1]_{ij}^{AB}=\epsilon^{AB} \gamma_{ij}$ introduced in Eq.~\eqref{eq:perturbed-nodewise}. For a given spin configuration $\{j_\ell\}$, we need to determine
\begin{multline}
\ket{\Gamma,\gamma_1,\{j_\ell\}} \equiv \frac{1}{\prod_n (J_n+1)!}\prod_{\ell} \frac{1}{[(2 j_\ell)!]^2}  \left(\delta^{AB} \delta^{CD} \frac{\partial^4}{\partial w^A_{s(\ell)}\partial\bar{w}^B_{s(\ell)} \partial w^C_{t(\ell)} \partial \bar{w}^D_{t(\ell)}}\right)^{2 j_\ell} \\
 \times \exp\left( \frac{1}{2}  \gamma_{ij} \epsilon^{AB} \bar{w}_A^i \bar{w}_B^j + \frac{1}{2} F_{ij}^\dagger \epsilon^{AB} w^i_A w^j_B\right) \ket{0} \Big|_{w=0} \, .
\end{multline}
For the moment, consider $\epsilon=0$. To evaluate the derivatives of the exponential, we must sum over all pairings of derivatives $\partial/\partial w$ ($\partial/\partial \bar{w}$) and complex variables $w$ ($\bar{w}$).  The $\partial/\partial \bar{w}$ derivatives pair off and pull down factors of $\epsilon^{AA'} \gamma_{ij}$. For a fixed pairing of $\partial/\partial \bar{w}$ derivatives, the $\epsilon^{AA'}$ tensors contract off with pairs of $\partial/\partial w$ derivatives, thus producing $D_{ij}$ operators. Since $\gamma_{ij}$ is block diagonal (for $\epsilon =0$), all of the $D_{ij}$ will be unmixed and can be evaluated with Proposition (\ref{deriv_formula}). We end up with products of the form 
\begin{equation}
(\gamma_{ij} \gamma_{i'j'}... )(F^\dagger_{ij} F^\dagger_{i'j'}...)
\label{eq:cycle-term}
\end{equation}
 where the $(i,j)$ pairs are determined by the specific pairing of the $\partial/\partial \bar{w}$ derivatives. We must then sum over all possible pairings.  These pairings are in one-to-one correspondence with routings of the graph. To see this, note that each pair $(i,j)$ in Eq.~\eqref{eq:cycle-term} corresponds to an oriented wedge. Furthermore, each fourth-order derivative in \eqref{eq:jl-coefficient} is naturally associated with a strand at a link $\ell$. Gluing such strands and wedges together, we obtain a routing $R$ with spin configuration $\{j_\ell\}$. The product \eqref{eq:cycle-term} then corresponds to $\gamma_\Phi F_\Phi^\dagger$, where $\Phi$ is the multiloop determined by $R$. Taking into account that there are $(2j_\ell)!$ routings $R$ in the equivalence class of $\Phi$, we find:
\be
\ket{\Gamma,\gamma_1,\{j_\ell\}} = \sum_{\Phi \in \{j_\ell\}}\frac{1}{\prod_{\ell} (2 j_\ell)!}  \gamma_\Phi F_\Phi^\dagger \ket{0} \, .
\ee
Summing over all spin configurations, we obtain \eqref{eq:cycle-decomp} with $\epsilon=0$.

We now take $\epsilon \ne 0$ and compute the first non-vanishing contribution. The process is the same as that described above, except that now we will have mixed $D_{ik}$ operators with $i$ and $k$ indices belonging to different nodes. Since $F^\dagger$ is block diagonal (even when $\gamma$ is not), $\partial/\partial w$ derivatives must pair off at nodes, and hence there must be an even number of total fibers associated with each node. This means that at minimum, two off-diagonal $\partial/\partial \bar{w}$ derivatives must pair off.\footnote{Suppose for instance there was only one off-diagonal $\partial/\partial \bar{w}$ pairing. Then we would be left with an odd number of $\partial/\partial \bar{w}$ at a single node, and hence cannot complete the pairing process \textit{within} the node. Instead, we must pair off the extra $\partial/\partial \bar{w}$ off-diagonally with another node.} Hence, the first non-vanishing off-diagonal correction is $O(\epsilon^2)$.

Furthermore, only two nodes will be involved in the off-diagonal pairing; three or more would take us to $O(\epsilon^3)$ or higher. Thus, we will have terms with exactly two mixed derivative operators $D_{ik}D_{j\ell}$  (with $i,j \in n$ and $k, \ell \in n'$ for $n\ne n'$) which we can evaluate using Proposition (\ref{deriv_disentangle}).  Since Proposition (\ref{deriv_disentangle}) yields unmixed derivatives, we obtain exactly the same terms as we did for $\epsilon =0$, except that we now have a factor $\gamma_{ik} \gamma_{j\ell}$. The factor involving $F^\dagger$ operators will remain identical to the $\epsilon=0$ case. We thus obtain a contribution of the form:
\begin{equation}
\frac{1}{2}(\gamma_{ik} \gamma_{j\ell}F^\dagger_{ij} F^\dagger_{k\ell}) \times (\text{other factors identical to the $\epsilon=0$ case}).
\end{equation}
Note that we will also have a derivative term like $D_{i\ell}D_{jk}$. This will result in
\begin{equation}
\frac{1}{2}(-\gamma_{i\ell} \gamma_{jk}F^\dagger_{ij} F^\dagger_{k\ell}) \times (\text{other factors identical to the $\epsilon=0$ case})
\end{equation}
where we have made use of $F^\dagger_{\ell k} = -F^\dagger_{k \ell}$. We can combine these contributions using
\be
\frac{1}{2}(\gamma_{ik} \gamma_{j\ell} - \gamma_{i\ell} \gamma_{jk})(F^\dagger_{ij} F^\dagger_{k\ell})=\frac{1}{2}B(\gamma_{ij} \gamma_{k\ell})(F^\dagger_{ij} F^\dagger_{k\ell}) \, ,
\ee
where $B(\cdot)$ was defined in \eqref{eq:braiding}. Finally, we must sum over all possible mixed derivative pairs, which, after extending the definition of $B(\cdot)$ as in \eqref{eq:braid-extend}, yields the result stated in \eqref{eq:cycle-decomp}.


%

\end{document}